\documentclass[%
 reprint,
superscriptaddress,
frontmatterverbose, 
 amsmath,amssymb,
 aps,
prb
floatfix
]{revtex4-2}
\usepackage{physics}
\usepackage{amsmath}
\usepackage{graphicx}
\usepackage{comment}
\usepackage{algorithm}
\usepackage{algorithmic}
\usepackage{tabularkv}
\usepackage[utf8]{inputenc}
\usepackage[toc,page,header]{appendix}
\usepackage[normalem]{ulem}
\usepackage[caption=false]{subfig}
\usepackage{bbm,braket,microtype,mathrsfs,amssymb,color,amsthm,mathtools,fullpage,enumitem,ae,aecompl,bm,thm-restate}
\usepackage[bookmarksnumbered,hypertexnames=false,colorlinks,colorlinks=true,linkcolor=blue,urlcolor=black,citecolor=blue,anchorcolor=green]{hyperref}
\begin{document}
\preprint{APS/123-QED}
	\title{Entanglement complexity of the Rokhsar-Kivelson-sign wavefunctions}

	%\author[1]{Stefano Piemontese}
    %\affil[1]{Physics Department,  University of Massachusetts Boston,  02125, USA}
    %\author[2]{Tommaso Roscilde}
    %\affil[2]{Univ Lyon, Ens de Lyon, CNRS, Laboratoire de Physique, F-69342 Lyon, France}
    %\author[1,3,4]{Alioscia Hamma}
    %\affil[3]{Dipartimento di Fisica ``Ettore Pancini'', Universit\`a degli Studi di Napoli Federico II,
    %Via Cintia, 80126 Napoli, Italy}
    %    \affil[4]{
    %INFN, Sezione di Napoli, Italy}
    
    \author{Stefano Piemontese}
    \email{s.piemontese001@umb.edu}
    \affiliation{Physics Department,  University of Massachusetts Boston,  02125, USA}

    \author{Tommaso Roscilde}
    \affiliation{Univ Lyon, Ens de Lyon, CNRS, Laboratoire de Physique, F-69342 Lyon, France}
    
    \author{Alioscia Hamma}
    \affiliation{Physics Department,  University of Massachusetts Boston,  02125, USA}
    \affiliation{Dipartimento di Fisica ``Ettore Pancini'', Universit\`a degli Studi di Napoli Federico II,
    Via Cintia, 80126 Napoli, Italy}
    \affiliation{INFN, Sezione di Napoli, 80126 Napoli, Italy}

	%\date{}

	%\section*{Abstract}
 \begin{abstract}
	Entanglement comes in different forms, some more complex than others. In this paper we study the transitions of entanglement complexity in an exemplary family of states---the Rokhsar-Kivelson-sign wavefunctions---whose degree of entanglement is controlled by a single parameter. This family of states is known to feature a transition between a phase exhibiting volume-law scaling of entanglement entropy and a phase with sub-extensive scaling of entanglement, reminiscent of the many-body-localization transition of disordered quantum Hamiltonians. We study the singularities of the Rokhsar-Kivelson-sign wavefunctions and their entanglement complexity across the transition using several tools from quantum information theory: fidelity metric, entanglement spectrum statistics, entanglement entropy fluctuations, stabilizer R\'enyi Entropy, % resources beyond the stabilized formalism;
	and the performance of a disentangling algorithm.  Across the whole volume-law phase the states feature universal entanglement spectrum statistics. Yet a "super-universal" regime appears for small values of the control parameter in which all metrics become independent of the parameter itself, the entanglement entropy as well as the stabilizer R\'enyi entropy appear to approach their theoretical maximum, the entanglement fluctuations scale to zero as in output states of random
	universal %non-Clifford 
	circuits, and the disentangling algorithm has essentially null efficiency. All these indicators consistently reveal a complex pattern of entanglement.  In the sub-volume-law phase, on the other hand, the entanglement spectrum statistics is no longer universal, entanglement fluctuations are larger and exhibiting a non-universal scaling, and the efficiency of the disentangling algorithm becomes finite. Our results, based on model wavefunctions, suggest that a similar combination of entanglement scaling properties and of entanglement complexity features may be found in high-energy Hamiltonian eigenstates---a very strong candidate being offered by the many-body localization transition of disordered lattice-spin models.

	%We construct an order parameter from the relative effective disentangling performance and propose an order-disorder quantum phase transition with universal critical indexes. 	 
	%\section{Introduction and analysis of the model}
	 \end{abstract}
  \maketitle
	\section{Introduction}
	
	\subsection{Entanglement and its complexity}
	Quantum entanglement is often considered as being the defining characteristic of quantum mechanics; and it is a fundamental notion for some of the behavior that is impossible for local classical physics, such as Bell nonlocality \cite{bell2004speakable, Scaranibook}.  In the context of quantum information,  quantum entanglement represents the resource behind the computational speed-up offered by quantum algorithms, and the central ingredient enabling protocols such as quantum teleportation \cite{bennett1993teleporting,Nielsen,shor1994algorithms,masanes2006all}.
	
	Since entanglement represents a fundamental form of quantum correlations between different local parts of a system, the question of its physical significance in quantum many-body systems has surged to great importance in the past two decades \cite{amico2008entanglement,eisert2010colloquium}. Quantum many-body systems are  typically governed by Hamiltonians featuring local interactions, and this locality is responsible for the spatial decay of correlations in the ground state; as well as for the area-law scaling of the entanglement entropy of a subsystem \cite{eisert2010colloquium,hamma2012quantum,hamma2012ensembles}, corresponding \emph{e.g.} to the von Neumann entropy of the subsystem density matrix. 
Entanglement has provided fundamental insights in the study of quantum many-body phenomena, such as quantum phase transitions (QPTs) \cite{Osterloh2002scaling, osborne2002entanglement, vidal2003entanglement,wu2004quantum, roscilde2004studying,hofmann2014scaling,amico2009entanglementmagnetic}; exotic phases of matter featuring topological order \cite{hamma2005ground,hamma2008entanglement,kitaev2006topological,Wen2007quantum,castelnuovo2008quantum}; and quantum many-body localization \cite{bardarson2012unbounded,vosk2013many-body,serbyn2013universal,serbyn2013local,serbyn2014interferometric,chen2015mbl,abanin2019colloquium}. 

More recently, it has been recognized that entanglement properties of wavefunctions go beyond a single quantity such as the entanglement entropy of a subsystem. Beyond the notion of bipartite and multipartite entanglement \cite{szalay2012partial,walter2016multipartite,bengtsson2017multipartite}, it has also been recognized that entanglement can be more or less \emph{complex} \cite{chamon2014emergent,Shaffer2014irreversibility,Roberts2017chaos}, in relationship with the resources needed to produce it. The complexity of entanglement can be associated with an ensemble of properties that appear to be concomitant, namely: 
\begin{itemize}
\item the emergence of a universal entanglement spectrum statistics,  namely a universal distribution for the gaps of the entanglement spectrum, compatible with the Wigner-Dyson distribution; 
\item the fact that it can only be produced by resources that go beyond those required to produce stabilizer states, 
%%%%%
namely quantum gates that fall outside of the Clifford group and are sufficient for universal quantum computation \cite{boykin2000anew,gottesman2009anintroduction}, 
%%%%%
 which are responsible for any computational advantage in quantum information processing \cite{Gottesman:1998hu}. This notion of ``non-stabilizerness" has also been commonly referred to  as {\em magic} in the recent literature \cite{bravyi2005universal, Knill2005quantum,Veitch2014resource,howard2017application, leone2022stabilizer}. 
\item the scaling of entanglement fluctuations corresponding to those of random states in the Hilbert space,  as well as of output states of random circuits using %non-Clifford
universal resources \cite{Leone2021quantum, true2022transitions}.
\item the failure of a disentangling algorithm based on Metropolis sampling  of Clifford random circuits~\cite{chamon2014emergent,Shaffer2014irreversibility, true2022transitions}, whereas states with simple patterns of entanglement can be efficiently disentangled by such annealing methods.
\end{itemize}

Interestingly, the presence of magic is in principle independent of the scaling of the entanglement entropy. Indeed there are states which are not entangled and which possess magic -- \emph{e.g.} states produced by single-qubit gates which are out of the Clifford group. 
Nonetheless, the complexity of entanglement  is generally thought of as stemming from the conjunction of a volume-law scaling for entanglement entropies, and a finite magic. Indeed these properties, along with a universal entanglement spectrum statistics, have been shown to be at the root of the onset of quantum chaos \cite{Leone2021quantum,leone2021isospectral,oliviero2021random,leone2022magic}; the hardness of disentangling algorithms \cite{chamon2014emergent,Shaffer2014irreversibility,true2022transitions}; the universal behavior of out-of-time-order correlation functions (OTOCs); \cite{kitaev2014hidden,Roberts2017chaos,harrow2021separation}; and the hardness of simulatability of quantum many-body systems \cite{Gottesman:1998hu,yang2017entanglement}.  
% the complexity of entanglement comes about from the conjunction of a volume law for entanglement and high values of the resource known as magic\cite{bravyi2005universal,Knill2005quantum}, which in turn is directly connected to classical simulability of quantum states through the Gottesman-Knill theorem\cite{Gottesman:1998hu}. 

While the role of entanglement properties in quantum many-body systems has now been firmly established \cite{amico2008entanglement,eisert2010colloquium}, it is only very recently that the connection between non-stabilizerness/magic and states produced by relevant quantum many-body physics has been the subject of investigation \cite{yang2017entanglement,liu2020many, oliviero2022magic}. Relevant progress in this direction has been made in Ref.~\cite{leone2022stabilizer}, where a novel measure of non-stabilizerness -- called stabilizer R\'enyi entropy -- has been introduced, which is experimentally measurable \cite{oliviero2022measuring} and directly amenable to a numerical computation for quantum many-body states \cite{oliviero2022magic, haug2023quantifying}. 
 
 The relationship between entanglement and non-stabilizerness/magic in physically relevant quantum many-body states has first been explored in the context of quantum circuits, where efficient entangling resources can be in principle limited to the Clifford group; while magic is only produced by universal 
 %non-Clifford
 operations \cite{chamon2014emergent,Shaffer2014irreversibility, zhou2020single}. Indeed it has been shown that  output states of random Clifford circuits possess volume law for entanglement just like Haar-random states; however, their entanglement spectrum statistics is not Wigner-Dyson distributed, but rather Poissonian. Moreover, the onset of Wigner-Dyson statistics in the output states of random universal%non-Clifford 
 circuits is accompanied by the appearance of  universal scaling in the fluctuations of entanglement entropy, while the fluctuations of output states of random Clifford circuits display non-universal scaling \cite{Leone2021quantum,oliviero2021transitions,true2022transitions} %{\bf (MAYBE WE SHOULD ADD THE FLUCTUATIONS OF ENTANGLEMENT AS A THIRD CHARACTERISTIC OF COMPLEXITY? WE USE IT NOW FOR THE CHARACTERIZATION OF THE TWO PHASES!).}
 ``Simple" entanglement, associated with the absence of magic, has  another relevant algorithmic consequence: an entanglement annealing algorithm, which generates a disentangling random Clifford circuit via the Metropolis algorithm, is very efficient in disentangling a state featuring simple entanglement, without any information on the circuit that generated it in the first place. 
 %From the quantum machine learning point of view, this corresponds to the learning of the circuit that produced the state \cite{zhou2020single} {\bf (IS THIS SENTENCE ADDING ANY INFORMATION? -- OR MAYBE I AM ALLERGIC TO STATEMENTS ABOUT MACHINE LEARNING...)}. This kind of quantum machine learning process is also at the root of the unscrambling of information from Hawking radiation in black holes modeled by random quantum circuits \cite{leone2022tolearn}.
  On the other hand, by gradually ``doping" random Clifford circuits with universal
  %non-Clifford 
  operations, such as T-gates, one can drive a transition towards a complex pattern of entanglement \cite{Leone2021quantum,oliviero2021transitions,true2022transitions,leone2022retrieving,leone2022learning,oliviero2022blackhole}.
 
  In the case of quantum many-body states which are eigenstates of a local Hamiltonian, entanglement complexity exhibits a very rich palette of possibilities. Disordered Heisenberg spin chains, corresponding to interacting 1d fermions with a random chemical potential, and featuring a many-body localization transition, have been shown \cite{yang2017entanglement} to exhibit two different forms of entanglement spectrum statistics. At low disorder, the model  possess high-energy eigenstates obeying the eigenstate-thermalization hypothesis (ETH) \cite{mahler2005emergence,popescu2006entanglement,goldstein2006canonical,reimann2007typicality,rigol2008thermalization}, and exhibiting a volume-law scaling for entanglement entropies. These states are also shown to feature a universal Wigner-Dyson entanglement spectrum statistics; and for these states the disentangling algorithm fails. Therefore the ETH phase corresponds to a complex-entanglement phase. Upon increasing the disorder strength, the many-body localization transition is accompanied by a violation of the Wigner-Dyson entanglement spectrum statistics, and by area-law scaling of entanglement. As a result, the disentangling algorithm can achieve a significant reduction of the entanglement entropy via a random Clifford circuit, although the convergence of this algorithm for such states has not been studied. A third form of entanglement complexity is exhibited by disordered XX spin chains, corresponding to non-interacting fermions with a random chemical potential. For those systems, all states feature Anderson localization for any disorder strength, namely an area-law scaling of entanglement; and concomitantly they show a Poisson entanglement spectrum statistics and are efficiently disentangled by an annealed random Clifford circuit~\cite{yang2017entanglement}.  
 
 A different picture is found instead in the ground state of paradigmatic local Hamiltonians such as the quantum Ising model: there entanglement is known to exhibit an area-law scaling, translating into a  a half-system entanglement entropy which is $O(1)$ in one dimension for every point of the phase diagram, except at the critical point of the ferromagnetic-paramagnetic transition \cite{vidal2003entanglement}. However, the stabilizer entropy exhibits {\em always} a volume-law scaling. In such models, the critical point is special because it is the point where magic also delocalizes and cannot be resolved in terms of local quantities \cite{oliviero2022magic}: in the gapped phase of a chain with $N$ qubits the magic $M_L$ of a block of $L$ spins would serve as a good approximation for  the magic of the whole chain $M_N$ by $M_N\simeq N/L M_L$. On the other hand, at the critical point, long-range correlations manifest themselves by an error in the approximation scaling with $L^{-1}$.
 
The ground state of the 1d quantum Ising model does not  exhibit a complex entanglement pattern, as its entanglement spectrum statistics is far from the Wigner-Dyson distribution. Indeed, it features also other symptoms of simple entanglement pattern: it is efficiently disentangled by entanglement annealing\cite{chamon2014emergent,true2022transitions},  lacks a volume-law scaling for entanglement entropy in spite of the volume-law scaling for stabilizer entropy. This is likely going to be the case for the ground states of most local Hamiltonian, given that an area-law scaling of entanglement (up to logarithmic corrections) is their characteristic feature. %{ {\bf I WOULD OMIT THE FOLLOWING SENTENCE, IT IS A BIT OBSCURE AT THIS STAGE!} \sout{However, the presence of long-range correlations like those induced by frustration may show up in more complicated patterns of both entanglement and magic\cite{inpreparation}.}} 
Moreover, even in the presence of  volume-law entanglement, a volume-law scaling for the stabilizer R\'enyi does not imply universal entanglement spectrum statistics,  but rather this scaling should be faster than a specific threshold as shown in Refs.~\cite{Leone2021quantum,oliviero2021transitions,true2022transitions}.

The unifying picture emerging from these results is the following. Complex entanglement, exhibiting volume-law scaling for entanglement entropy {\em and} a Wigner-Dyson entanglement spectrum statistics  is a characteristic of excited Hamiltonian eigenstates satisfying ETH, as well as states prepared with random universal %non-Clifford 
circuits. On the other hand, entangled states which are not complex %-- as they lack either a volume-law scaling of entanglement entropy, or Wigner-Dyson ESS, or extensive stabilizer R\'enyi entropy, or {\bf (WHAT ELSE?)} 
are found as eigenstates of strongly disordered Hamiltonians, as output states of random Clifford circuits, and ground states of local Hamiltonians.

 \begin{figure}
		%\centering
		\includegraphics[clip,width=\columnwidth]{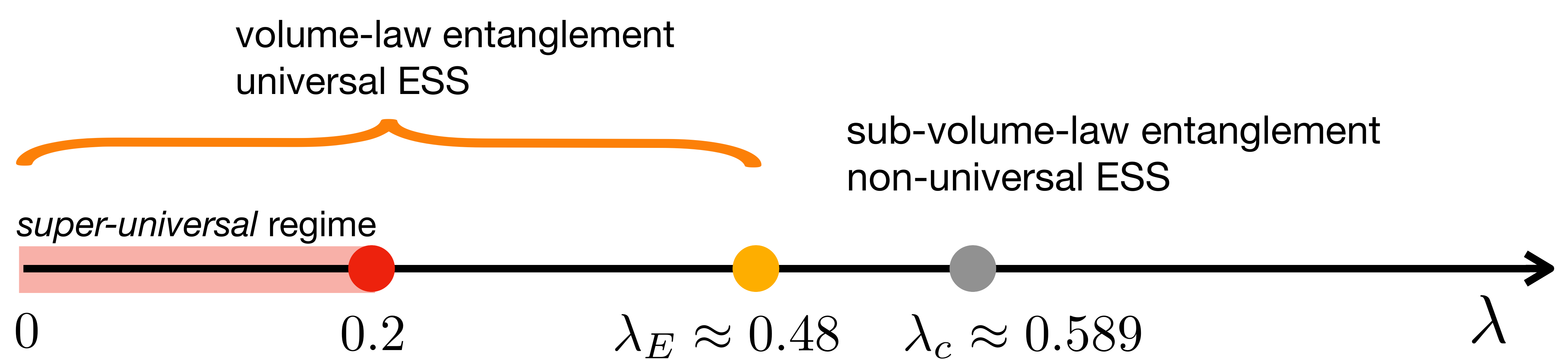}
%		\subfloat[]{\includegraphics[clip,width=0.5\columnwidth]{images/fidelity/fidelity_fit.png}\label{fig_fidelity_metric_b}}
		\caption{
		Schematic diagram of the various entanglement phases we found for the RK-sign states. The entanglement transition at $\lambda_E$ divides a phase with volume-law scaling of the entanglement entropy and universal (Wigner-Dyson) entanglement entropy statistics (ESS), from a phase with sub-volume-law entropies and non-universal ESS. A further super-universal regime is identified within the volume-law phase. 
%		b) Polynomial fit of the fidelity metric curves depicted in a), derived using the function \emph{polyfit()} from the Python \emph{Numpy} library\cite{harris2020array}. Computing the derivatives of these polynomials allows us to derive the exact position of the critical point in the thermodynamic limit.
		}\label{phase_diagram}
	\end{figure}

\subsection{Probing entanglement complexity in model wavefunctions}

The above emerging picture is so far only reported ``piecewise" in the recent literature -- namely only partial aspects of entanglement complexity have been investigated for either Hamiltonian eigenstates or output states of random circuits. In this paper we aim at a thorough investigation of entanglement complexity in a family of quantum many-body states which have the property of admitting an explicit expression for their coefficients, depending on a single parameter; and which feature a  rich palette of different phases with varying entanglement complexity upon tuning the control parameter. This one-parameter family of states has the so-called Rokhsar-Kivelson (RK) form, equipped with a random sign structure (hereafter denoted as RK-sign family) -- namely the wavefunction coefficients in the computational basis have amplitudes given by the Boltzmann weights of the so-called random-energy model  (REM) \cite{derrida1980rem} of statistical mechanics, and random signs.   In particular, the RK model is based on a quantum random energy model (QREM). This model can be seen as the limit $p\to \infty$ of the $p-$spin model (see \cite{manai2020phase}). As such, the RK model is non-local.

The parameter $\lambda$ controlling the property of the states corresponds to the inverse temperature for the REM. These states have been first studied in Ref.~\cite{chen2015mbl}, where they have been found to exhibit a transition from a phase possessing volume-law scaling of entanglement, to a phase featuring sub-extensive scaling, reminiscent of many-body localization.

%stabilizerness that are typical of the ETH phase, while a simple pattern of entanglement is established by low-entanglement and  below threshold SRE. It seems thus that the MBL transition from ETH to the localized phase is paradigmatic to describe the onset (or demise) of complex entanglement in quantum many-body states. 
% It would be thus important to study a quantum many-body model where the volume law is established at the same time for both entanglement and non-stabilizerness in one complex entangled phase, and transitions to a non-volume law for both resources in the phase governed by a simple pattern of entanglement.
%In this paper, we study  the complexity of entanglement  in a one parameter$-\lambda$ family of states of the Rokhsar-Kivelson (RK) form equipped with a random sign structure\cite{grover2014quantum,grover2015entanglement}. These states have been studied in\cite{chen2015mbl} to show a transition from a volume law, ETH phase, to many-body localized phase featuring entanglement area law.  The two phases are separated by a critical point $\lambda_c$ revealed by the fidelity  metric. This phase transition corresponds to the classical phase transition of the random energy model\cite{derrida1980rem}.  

In this work, we show rigorously that the RK-sign states possess a transition at a critical value $\lambda_c$ revealed by the fidelity metric, and corresponding to the spin-glass phase transition of the REM. Nonetheless the existence of this transition is in fact independent of the sign structure of the states, while the entanglement properties are crucially dependent on such a structure. As a consequence the entanglement properties reveal a different picture, exhibiting a rich palette of different scaling regimes and complexity features. We investigate all the known complexity metrics for entanglement, namely: (i) the scaling of the half-system entanglement entropy; (ii) the adherence of the entanglement spectrum statistics to the Wigner-Dyson distribution, as measured by the Kullback-Leibler divergence $D_{\rm KL}$, and by the average ratio of adjacent gaps in the entanglement spectrum; (iii) the amount of non-stabilizerness measured by the stabilizer R\'enyi entropy $M_2$; (iv) the  behavior of entanglement ensemble fluctuations; and (v) the hardness of factorizing by means of an entanglement annealing algorithm \cite{chamon2014emergent,Shaffer2014irreversibility,true2022transitions}. 

In agreement with Ref.~\cite{chen2015mbl} we find that the RK-sign states possess two scaling regimes for the half-system entanglement entropy, namely a volume-law scaling for $\lambda < \lambda_E$ and a sub-volume-law scaling for $\lambda > \lambda_E$, with $\lambda_E < \lambda_c$. The volume-law regime is characterized by the adherence of the entanglement spectrum to the Wigner-Dyson statistics, and as such it can be qualified as featuring complex entanglement; on the other hand, the sub-volume-law phase exhibits a non-universal entanglement spectrum statistics. Nonetheless for the whole range of parameters, encompassing the two above-cited phases, the stabilizer R\'enyi entropy is extensive, revealing that the entanglement of the RK states requires universal
%non-Clifford 
resources to be prepared. Most remarkably, within the volume-law phase a \emph{super-universal} regime exists for $\lambda \lesssim 0.2$, in which all complexity metrics become essentially independent of $\lambda$ and take a universal behavior: 
%the entanglement entropy approaches in scaling the Page entropy of random states \cite{Page1993}; 
the stabilizer R\'enyi entropy scales to its maximum possible value; the fluctuations of entanglement entropy exhibit a universal scaling to zero as in random quantum states; and the disentangling algorithm fails completely to reduce the entanglement entropy of the state. A sketch of the various entanglement regimes is given in Fig.~\ref{phase_diagram}.

This paper is organized as follows. In Sec.~\ref{s.RKsign} we introduce the RK-sign family, and  show the existence of a singularity in the fidelity metric. In Sec.~\ref{s.complexity} we describe our results for the entanglement entropy scaling and its ensemble fluctuations, the entanglement spectrum statistics, the stabilizer R\'enyi entropy and the efficiency of a disentangling algorithm. Conclusions and perspectives are offered in Sec.~\ref{s.discussion}.

	\section{Quantum Phase Transition of the Rokhsar-Kivelson-sign family}
	\label{s.RKsign}
	
\subsection{The RK-sign wavefunctions}
	
In this section, we define the family of states dubbed RK-sign, and study their fidelity metric \cite{quan2006decay,zanardi2006ground,abasto2008fidelity}.
We consider a system of $N$ qubits, whose Hilbert space is spanned by the computational basis $|{\bm \sigma} \rangle  = |\sigma_1, \sigma_2, ... , \sigma_N\rangle$ of eigenstates of the $\sigma^z$ Pauli operators. The RK-sign family \cite{castelnovo2005from,chen2015mbl}, dependent on the parameter $\lambda$, is then defined as
	\begin{equation}\label{rksign_model}
		\ket{\psi(\lambda)} = \frac{1}{\sqrt{Z(\lambda)}}\sum_{{\bm \sigma}}W_{\bm \sigma}~ e^{-%\frac
			{\lambda}%{2}
			E_{\bm \sigma}}\ket{{\bm \sigma}}.
	\end{equation}
Here the  `energies' $E_{\bm \sigma}$ obey the so-called Random Energy Model (REM) \cite{derrida1980rem}
% laumann2014many-body,baldwin2016many-body} 	
namely they are normally distributed variables with zero mean and variance ${\rm Var}(E_{\bm \sigma}) = N$; the signs $W_{\bm \sigma}$ are random with probabilities $p(W_{\bm \sigma} = \pm 1) = \frac{1}{2}$. The normalization factor contains the partition function $Z(\lambda) = \sum_{\bm \sigma} e^{-2\lambda E_{\bm \sigma}}$, namely the partition function of the REM model at inverse temperature $T^{-1} = 2\lambda$ (taking $k_B=1$).  

%so that  $E_{\bm \sigma}$ and $W_{\bm \sigma}$ are extracted with probabilities
%	\begin{align}\label{eq_distrib}
%		p(W_{\bm \sigma} = \pm 1) &= \frac{1}{2}, \\
%		p(E_{\bm \sigma} = x) &= Gauss(x, \mu = 0, Var = \sqrt{N}). %'sigma' symbol used for configurations
%	\end{align}	
%Here $Z(\lambda)$ is given by $Z(\lambda) = \sum_{\bm \sigma} e^{-2\lambda E_{\bm \sigma}}$ and the 
%	$W_{\bm \sigma}=\pm1 $ are random signs. The `energies' $E_{\bm \sigma}$ obey the so-called Random Energy Model (REM) \cite{derrida1980rem, laumann2014many-body,baldwin2016many-body} %the last 2 ref. actually show qpt for the rem
%	equipped with a random sign structure\cite{grover2014quantum,grover2015entanglement,chen2015mbl}

% With the random sign $W_{\bm \sigma} =\pm 1$, this model has been  studied in\cite{chen2015mbl} to show a transition from a thermalized phase with volume law for entanglement, to an MBL phase. At a critical value $\lambda_{sg}= \sqrt{2\ln 2}/2\simeq .59$, the free energy density computed from the classical Sherrington-Kirkpatrick spin glass model is shown to be non-analytic because of a spin glass transition. For the RK-sign model, the authors in\cite{chen2015mbl} estimate an MBL phase transition fro $\lambda\le  \lambda_{sg}= \sqrt{2\ln 2}/2$.

For $W_{\bm \sigma} =1$, the above wavefunctions correspond to the so-called RK states, which are related to ground states of strongly frustrated quantum antiferromagnets \cite{rokhsar1988superconductivity}. On the other hand, the randomization of the sign is a necessary condition for the RK states to mimic the physics of excited eigenstates of local Hamiltonians  \cite{grover2014quantum,grover2015entanglement,chen2015mbl}, which exhibit in general a volume-law scaling of entanglement entropies. Indeed states which have random but equally signed coefficients are expected to not exhibit any scaling of the subsystem entanglement entropy \cite{grover2014quantum}. Hence, in order to represent states which have entanglement complexity, a crucial ingredient of the model is the addition of random signs, potentially leading to volume-law scaling of entanglement 
\cite{grover2014quantum,grover2015entanglement,chen2015mbl}. 

The scaling of entanglement entropy with subsystem size in the RK-sign states  is fundamentally controlled by the parameter $\lambda$, which governs the distribution of the amplitudes for the wavefunction coefficients.  Entanglement entropies are found to exhibit a transition from a volume-law scaling at small $\lambda$ to sub-volume-law scaling at large $\lambda$
\cite{chen2015mbl}. One may suspect that this transition is connected with the thermodynamic transition happening for the REM at  $\lambda_c= \sqrt{2\ln 2}/2\simeq 0.589$, at which the second derivative (the specific heat) and higher derivatives of the partition function $Z(\lambda)$ become singular. Yet, examining the scaling of R\'enyi entropies with different indices $n$ for a (1/3)-(2/3) bipartition of the system, Ref.~\cite{chen2015mbl} concludes that the entanglement transition occurs at smaller values of $\lambda$,  and in a way which is dependent on the R\'enyi index. 
 In Sec.~\ref{subs.entropy} we shall confirm this result by focusing on the von Neumann entropy (corresponding to $n=1$) for a (1/2)-(1/2) bipartition.
Nonetheless, in the remainder of this section we shall instead focus on another indicator of transitions in one-parameter families of wavefuctions, namely the fidelity metric, which can be proven rigorously to exhibit a transition at $\lambda_c$. 

\subsection{Singularity in the fidelity metric}
\label{subs.fidelity}

%	For this reason, all the different quantities that we introduce have to be averaged over multiple trials. The exact number of samples $N_s$ has to be estimated.

One-parameter families of quantum states, similar to the RK-sign one, can be easily obtained as eigenstates of a quantum Hamiltonian ${\cal H}(\lambda)$ dependent on a parameter. Singularities in such families of states as a function of $\lambda$ correspond therefore to (excited-state) quantum phase transitions of the Hamiltonian of interest. Even though we do not have a ``parent" Hamiltonian for the RK-sign family, it is simple to build one mathematically (via projectors onto the states of interest); and the random nature of the RK-sign states implies that the parent Hamiltonian is a disordered one, with disorder strength controlled by $\lambda$. The transition in the RK-sign model can therefore be viewed as a disorder-induced transition in excited eigenstates of a random Hamiltonian, akin to the many-body localization transition of disordered spin models.  

In order to investigate the existence of quantum phase transitions as a function of the parameter $\lambda$ of the RK-sign family, one can adopt a general strategy to detect the existence of quantum phase transitions in one-parameter families of states as \emph{singularities} in the dependence of the state properties on the parameter itself. Such a strategy is based on the behavior of the fidelity \cite{zanardi2006ground, gu2010fidelity}
between two nearby states parametrized by $\lambda_1\simeq \lambda_2$, defined as
$f(\lambda_1,\lambda_2) = \abs{\braket{\psi(\lambda_1)|\psi(\lambda_2)}}^2$. 
For every $\lambda$,  we can evaluate $f(\lambda,\lambda+\epsilon)$ for an infinitesimal $\epsilon$, allowing for the expansion $f(\lambda,\lambda+\epsilon) \approx 1 - g_{\lambda\lambda}~\epsilon^2 + o(\epsilon^3)$. The coefficient $g_{\lambda\lambda}$ is the so-called \emph{fidelity (quantum Fisher) metric}  \cite{zanardi2007information, abasto2008fidelity} or \emph{fidelity susceptibility} \cite{you2007fidelity}, 
 \begin{align}\label{fidelity_metric}
     g_{\lambda\lambda} &= \braket{\partial_\lambda\psi|\partial_\lambda\psi} - \abs{\braket{\psi|\partial_\lambda\psi}}^2 \\
     &= -\frac{1}{2}\frac{\partial^2}{\partial\epsilon^2}f(\lambda,\lambda+\epsilon)\bigg\rvert_{\epsilon=0}~,
     \end{align}
providing a notion of distance in Hilbert space among states belonging to the one-parameter family. 

Quantum phase transitions can be generically detected by a non-analicity~\cite{jacobson2009scaling} in $g_{\lambda\lambda}$, as shown in a wide variety of recent numerical studies \cite{zanardi2006ground, Zhou_2008, gu2010fidelity,Damski_2015}. Moreover the fidelity metric is the quantum Fisher information of the quantum state, expressing the fundamental sensitivity of the state to variations of the $\lambda$ parameter, and bounding the precision with which the $\lambda$ parameter can be estimated by making measurement on the state. The singularity of the fidelity metric at quantum phase transitions is therefore at the root of the enhanced metrological sensitivity for parameter estimation associated with quantum criticality \cite{Zanardietal2008}.

The fidelity metric for the RK-sign wavefunctions  can be easily computed analytically, and it can be directly related to thermodynamic properties of the REM -- related to the amplitudes of the wavefunction coefficients. We start by observing that the $\lambda-$derivative of the state vector $\ket{\partial_\lambda\psi}$ takes the form:
		\begin{align}
		    \ket{\partial_\lambda\psi} = \frac{1}{\sqrt{Z}}\sum_{\bm \sigma}\big(\langle E \rangle_{\rm REM} - E_{\bm \sigma}\big)W_{\bm \sigma} e^{-\lambda E_{\bm \sigma}}\ket{{\bm \sigma}}.
		\end{align}
		where $\langle E^n \rangle_{\rm REM} \equiv  \frac{1}{Z}\sum_{\bm \sigma} E^n_{\bm \sigma} e^{-2\lambda E_{\bm \sigma}}$ is the thermodynamic average for the REM. 
		This allows us to compute the two terms appearing in the expression of the fidelity metric, Eq.~\eqref{fidelity_metric}:
		\begin{align}
		    \braket{\partial_\lambda\psi|\partial_\lambda\psi} &= \frac{1}{Z}\sum_{\bm \sigma}\big(\langle E\rangle_{\rm REM} - E_{\bm \sigma}\big)^2 e^{-2\lambda E_{\bm \sigma}} \\
%		    &= \frac{1}{Z}\sum_{\bm \sigma}\big(\langle E\rangle^2 - 2\langle E\rangle E_{\bm \sigma} + E^2_{\bm \sigma}\big)e^{-2\lambda E_{\bm \sigma}} \\
		    %&= \langle E\rangle^2 - 2\langle E\rangle^2 + \langle E^2\rangle \equiv \Delta^2 E \\
		    &= \langle E\rangle_{\rm REM}^2 - \langle E\rangle_{\rm REM}^2 \\
		    \braket{\psi|\partial_\lambda\psi} &= \frac{1}{Z}\sum_{\bm \sigma} \big(\langle E\rangle_{\rm REM} - E_{\bm \sigma}\big)e^{-2\lambda E_{\bm \sigma}} %\\ 
		    %&= \langle E\rangle \frac{\sum_{\bm \sigma} e^{-2\lambda E_{\bm \sigma}}}{Z} - \langle E\rangle 
		    = 0.
		\end{align}
		Hence the fidelity metric, averaged over the ensemble of energies $\{E_{\bm \sigma} \}$, is equal to the energy fluctuations of the REM at inverse temperature $T^{-1} = 2\lambda$:
		\begin{equation}\label{eq_g_fluct}
			\overline{ g_{\lambda\lambda} }= \langle E^2\rangle_{\rm REM}^2 \equiv \left (\Delta^2 E \right)_{\rm REM}~
		\end{equation}
	where $\overline{(...)}$ denotes the average over the ensemble of RK-sign states at fixed $\lambda$.	
		
	The energy fluctuations of the REM are directly related to its heat capacity, via the fluctuation-dissipation relation
	\begin{equation}
	C_{\rm REM} = \frac{\partial \langle E\rangle_{\rm REM}}{\partial T} =  \frac{\left (\Delta^2 E \right)_{\rm REM}}{T^2}~.
	\end{equation}
	In turn the heat capacity of the REM is analytically known \cite{derrida1980rem,derrida1980therem} to exhibit a jump singularity at the transition of the model:
	\begin{equation}
	C = \begin{cases} 0 ~~~~ T<T_c \\
	\frac{N}{T^2} ~~~~ T > T_c
	\end{cases}
	\end{equation}
	with $T_c = 1/\sqrt{2\ln{2}}$~.~\footnote{Notice that the critical temperature quoted here differs by a factor of 2 with respect to that of Ref.~\cite{derrida1980rem}, because the variance of energy fluctuations is ${\rm Var}(E_{\bm\sigma})= N/2$ in that work and it is ${\rm Var}(E_{\bm\sigma})= N$ in this work.} This immediately implies that the fidelity metric of the RK-sign family $\overline{g_{\lambda\lambda}} = T^2 C_{\rm REM}$ inherits the singularity of the specific heat of the REM model at $\lambda_c = 1/(2T_c) = \sqrt{2\ln 2}/2$ in the form of a simple step-function singularity: 
\begin{equation}
	\overline{g_{\lambda\lambda}} = \begin{cases} 0 ~~~~ \lambda > \lambda_c \\
	N ~~~~ \lambda < \lambda_c~.
	\end{cases}
	\end{equation}

  \begin{figure}
		\centering
		\includegraphics[clip,width=\columnwidth]{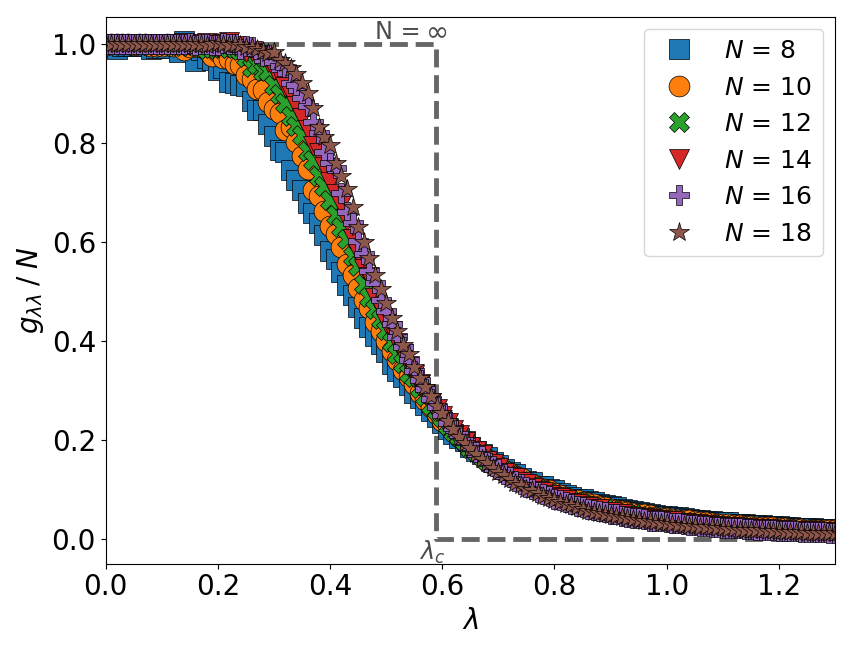}\label{fig_fidelity_metric_a}
%		\subfloat[]{\includegraphics[clip,width=0.5\columnwidth]{images/fidelity/fidelity_fit.png}\label{fig_fidelity_metric_b}}
		\caption{
		Plot of the fidelity metric $g_{\lambda\lambda}$ per qubit for different system sizes. Each point is an average over $N_s = 1000$ realizations of the wavefunction. The dashed line shows the expected behavior in the thermodynamic limit.
%		b) Polynomial fit of the fidelity metric curves depicted in a), derived using the function \emph{polyfit()} from the Python \emph{Numpy} library\cite{harris2020array}. Computing the derivatives of these polynomials allows us to derive the exact position of the critical point in the thermodynamic limit.
		}
		\label{fig_fidelity_metric}
	\end{figure}     
             %Numerically, this second derivative is computed via difference quotient with a small fixed $\epsilon$:
             %\begin{align}
             %\frac{\partial^2}{\partial{\lambda_2}^2}f(\lambda_1,\lambda_2) &= \frac{\partial}{\partial\lambda_2}\left(\frac{f(\lambda_1,\lambda_2+\epsilon)-f(\lambda_1,\lambda_2)}{\epsilon}\right)  \\
             %&=\frac{\frac{f(\lambda_1,\lambda_2+\epsilon+\epsilon')-f(\lambda_1,\lambda_2+\epsilon)}{\epsilon'}-\frac{f(\lambda_1,\lambda_2+\epsilon')-f(\lambda_1,\lambda_2)}{\epsilon'}}{\epsilon} \\
             %&= \frac{f(\lambda_1,\lambda_2 + 2\epsilon) - 2  f(\lambda_1,\lambda_2 + \epsilon) + f(\lambda_1,\lambda_2) }{\epsilon^2}.
             %\end{align}

\begin{figure}[H]
		\centering
		\subfloat[]{\includegraphics[clip,width=\columnwidth]{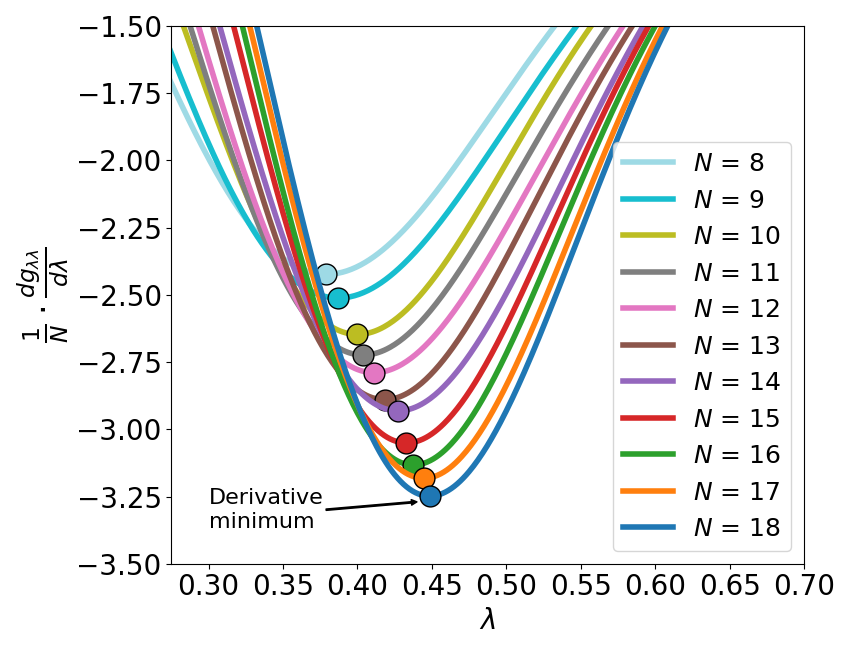}\label{fig_g_crit_a}}
		
  \subfloat[]{\includegraphics[clip,width=\columnwidth]{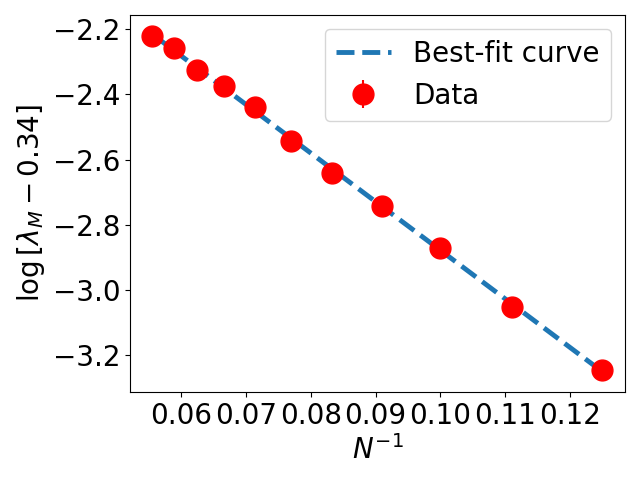}\label{fig_g_crit_b}}
		\caption{(a) First derivatives of the polynomials used to fit $g_{\lambda\lambda} / N$, for different system sizes $N$: the dots mark the minima $\lambda_M(N)$ of each curve. 
		(b) Localization of the critical point $\lambda_c$ through finite-size scaling: %{\color{blue}
  the minima $\lambda_M(N)$ are fitted to an exponential curve of the form $\lambda_M(N) = A \exp\left[\frac{-B}{N}\right] + C$. A $\chi^2$-minimization approach \cite{newville2014lmfit} gives $A=0.249\pm0.005$, $B=14.9\pm1.4$, $C=0.339\pm0.007$, and $\lambda_c$ is extrapolated to infinite size as $N^{-1} \rightarrow 0$. The results are here plotted on a semi-logarithmic scale, where a linear trend ($R^2 \approx 0.9994$) emerges.%} 
  %the minima $\lambda_M(N)$ are here plotted on a semi-logarithmic scale as a function of $N^{-1}$ and extrapolated to infinite size via an exponential fit (dashed curve). 
		}
		\label{fig_g_crit}
	\end{figure}

     The above analytical calculation is strictly valid only in the thermodynamic limit $N\to \infty$. It is instructive to examine finite-size effects by computing $g_{\lambda\lambda}$ numerically on finite-$N$ wavefunctions via the finite-element derivative formula: 
    \begin{equation}
         g_{\lambda\lambda} = -\frac{1}{2}\frac{f(\lambda,\lambda+2\epsilon)-2f(\lambda,\lambda+\epsilon)+1}{\epsilon^2}.
         \end{equation}
     For every realization of $W_{\bm \sigma}, E_{\bm \sigma}$, we compute $g_{\lambda\lambda}$ and average it over $N_s= 1000$ realizations. In Fig.~\ref{fig_fidelity_metric} we show the behavior of $\overline{g_{\lambda \lambda}}$ for different system sizes $N=8,\ldots, 18$. The curves of $\overline{g_{\lambda \lambda}}/N$ for different values of $N$ are found to cross in the vicinity of the critical value $\lambda_c$, consistent with the existence of a jump in the thermodynamic limit. 
  To reconstruct the position of the critical point $\lambda_c$ from finite-size scaling, we analyze the derivatives of $g_{\lambda\lambda}$ for our finite systems. For each $N$, we consider the polynomial that fits the points of $g_{\lambda\lambda}$ as a function of $\lambda$. Then, we take the first derivative of these polynomials and identify the minimum points $\lambda_M(N)$ (Figure \ref{fig_g_crit_a}). %{\color{blue} 
  We fit the minima $\lambda_M(N)$ to an exponential curve of the form $A\exp(-B/N)+C$, such that $\lambda_M(N\to \infty) = A+C$. This scaling law is phenomenological, but it is strongly suggested by our data (as shown in Fig.~\ref{fig_g_crit_b});  and the extrapolated value of $\lambda_M$, $\lambda_M(\infty) = 0.588 \pm 0.013$, is in quantitative agreement with the REM transition point $\lambda_c = \sqrt{2 \ln 2}/2 \simeq 0.589$. %The result of the fit gives 
  %{\color{blue}The choice of an exponential function solely comes from a numerical analysis (Fig.~\ref{fig_g_crit_b}) and} give. 
  This scaling analysis highlights the fact that finite-size effects are very strong for the system sizes we considered. %}
  % This may apply as well to the entanglement transition discussed below, suggesting that the accuracy of our estimates of the location of the transition might be severely affected by 

	\section{Entanglement Complexity Transition}
	\label{s.complexity}
	In this section, we examine the quantum phase transition of the RK-sign family from the point of view of entanglement complexity, computing the behavior of different complexity metrics as a function of the parameter $\lambda$. More precisely, we will compute the entanglement entropy for a subsystem with $N/2$ qubits, the entanglement spectrum statistics, the  stabilizer R\'enyi entropy, the ensemble fluctuations of the entanglement entropy, and the disentangling power of a Metropolis algorithm.

\begin{figure}
		\centering
		\subfloat[]{\includegraphics[clip,width=\columnwidth]{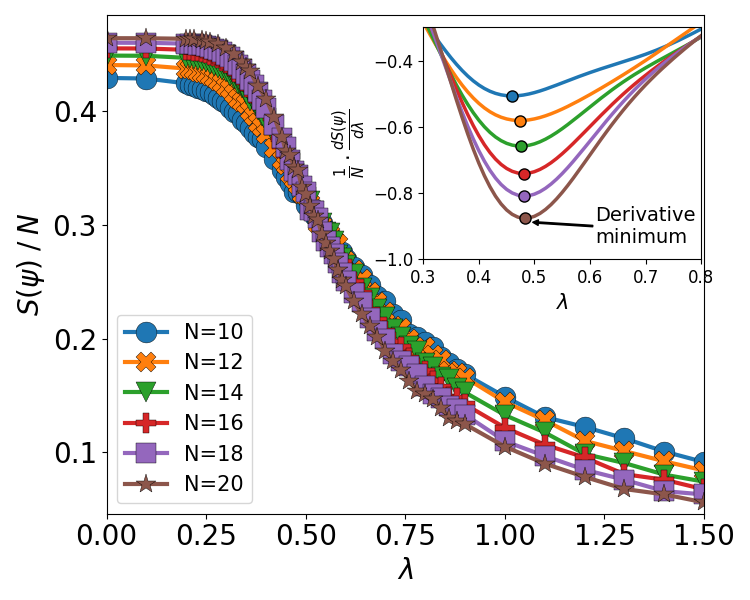}\label{fig_ent_ent_a}}
  
		\subfloat[]{\includegraphics[clip,width=\columnwidth]{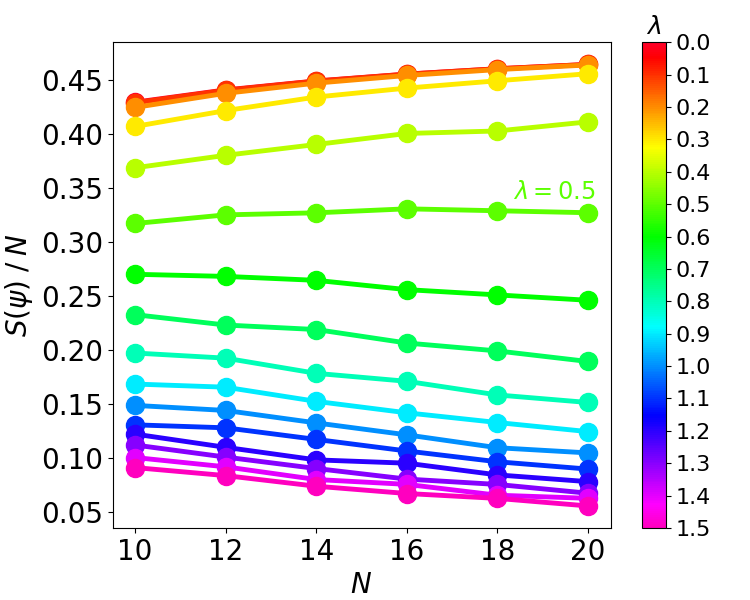}\label{fig_ent_ent_b}}
		\caption{ (a) Von Neumann Entanglement Entropy $S(\psi)$ per qubit as a function of $\lambda$, for different system sizes $N$. In the inset, first derivative of the best fit curves for $S(\lambda)$; the dots mark the minimum of each curve.  (b) $S(\psi) / N$, as a function of $N$, for different $\lambda$ in the interval $[0,1.5]$. 
		%{\bf ADD A SCALING ANALYSIS OF THE CROSSING POINTS, POSSIBLY SHOWING A DRIFT TOWARDS HIGHER VALUES -- OTHERWISE WE CANNOT CONCLUDE THAT THE TRANSITION IN THE SYSTEM IS UNIQUE...}
		}

		%  As expected, for low $\lambda$, the basis configurations have the same weight and the entanglement entropy is high, whereas the more  $\lambda$ is increased the more the system is dominated by a single configuration, leading to low values of $S(\psi)$. The crossing of the curves indicate that these two phases have a definite transition point in the thermodynamic limit. Such point appears to be compatible with the critical point ($\lambda_c \approx 0.588$) derived for the fidelity metric $g_{\lambda\lambda}$, here depicted with the dashed vertical line. 
	%	On the other hand, we see that for $\lambda\lesssim 0.2$ the entropy $S(\psi)(\lambda)$ is constant. In the following, this will be shown to correspond to the complex entangled phase. }
		\label{fig_ent_ent}
	\end{figure}

\subsection{Scaling of entanglement entropy}	
\label{subs.entropy}
	For every state $|\psi (\lambda)\rangle$ we consider an equal bipartition into two groups of $N/2$ qubits, and calculate the von Neumann entropy of the reduced density matrix of the half system, $S_{N/2}(\psi) = - {\rm Tr}(\rho_{N/2} \log_2 \rho_{N/2} ) $ where $\rho_{N/2} = {\rm Tr}_{N/2} |\psi\rangle \langle \psi |$. The half-system entropies are then averaged over a sample of size $N_s$ of RK-sign states, to give the results shown in Fig.~\ref{fig_ent_ent}. There we observe that, upon increasing $\lambda$, a clear transition takes place between a phase with  volume-law scaling of entanglement at low $\lambda$, and a phase with sub-extensive scaling. In fact the low-$\lambda$ phase appears to exhibit super-extensive scaling, but this can only be a finite-size effect, as extensive scaling is the fastest one allowed for entanglement entropy. Moreover, the volume-law phase remarkably contains a regime for $\lambda \lesssim 0.2$ where the entanglement entropy is constant, a feature that will be common to other figures of merit examined in this work. 
	
The location of the transition between the two scaling regimes  (volume law vs. sub-extensive scaling) can be in principle deduced  from the position of the minimum derivative of the $S_{N/2}/N$ curves: indeed, regardless of the specific scaling dimension of the entanglement entropy at the transition, one can generally expect that the critical point be marked by the strongest decrease of $S_{N/2}$ upon varying the control parameter. The inset of Fig.~\ref{fig_ent_ent_a} shows that the minimum derivatives exhibit little scaling in position, and indicate an entanglement transition at $\lambda_E \approx 0.48 < \lambda_c$. While this estimate might still be affected by significant finite-size corrections, our conclusion (entanglement transition preceding the REM transition) is in agreement with the findings of Ref.~\cite{chen2015mbl}. In fact it is easy to understand why the two transitions should be independent, based on rather general arguments.
As we have seen in Sec.~\ref{subs.fidelity}, the fidelity metric is uniquely sensitive to the \emph{amplitudes} of the wavefunction coefficients and their $\lambda$ dependence -- namely on the localized vs. delocalized nature of the wavefunctions when expressed on the computational basis. On the other hand, as shown in Ref.~\cite{grover2014quantum}, the sign structure of the coefficients is crucially important in determining the scaling of the entanglement entropy: if one considered \emph{e.g.} positive-definite coefficients, all RK states would exhibit a sub-volume-law scaling of entanglement entropy. This means that different sign distributions for the coefficients (less random than the one adopted here) can move the entanglement transition of the RK-sign family, or make it disappear altogether; while by construction the fidelity-metric transition remains insensitive to these features. Hence the entanglement transition and the REM transition are necessarily decoupled.  
%{\color{blue} 
This observation may look surprising, but in fact changes in the behavior of the entanglement pattern are not always accompanied by a singularity in the fidelity metric related to a QPT \cite{amico2006divergence,amico2009entanglementmagnetic}. As an example, Ref. \cite{tomasello2011ground} shows that the ground-state factorization occurring in the XYZ model in a field, albeit providing a special behavior in the entanglement properties, does not show any singularity in the fidelity metric related to the ground-state wavefunction.%}

\subsection{Entanglement spectrum statistics}	

  In this section we examine the entanglement spectrum statistics, which has been pointed out as a fundamental indicator of entanglement complexity in Refs.~\cite{chamon2014emergent,Shaffer2014irreversibility,zhou2020single}.
	Considering the eigenvalues of the half-system reduced density matrix $\{ a_i \}$ ($i = 1,\dots,2^{\frac{N}{2}} -1$) in increasing order $a_i < a_{i+1}$, 
	 we define the gaps between adjacent eigenvalues $\delta_j = a_{j+1} - a_j$, as well as the two different ratios of adjacent gaps:
	\begin{align}
		r_k &= \frac{\delta_{k+1}}{\delta_k}  \label{eq_R} \\
		\tilde{r}_k &= \frac{\min[{\delta_{k+1},\delta_k}]}{\max[{\delta_{k+1},\delta_k}]}~. \label{eq_R_tilde}
	\end{align}
	We then focus on the gap-ratio distribution $P(r)$ and on the average gap ratio $\langle \tilde{r} \rangle$, built from the gaps of all the density matrices associated with a sample of RK-sign states (of size $N_s=1000$) upon varying the $\lambda$ parameter.

%	The results of Refs.~\cite{Shaffer2014irreversibility,zhou2020single} indicate that entanglement complexity is associated with a specific form of entanglement spectrum statistics (ESS), namely the distribution of the gaps in the eigenvalues of the reduced density matrix. In particular, complex entanglement is associated with the appearance of Wigner-Dyson ESS.
 	
%	If the initial states is sampled a number $N_s$ of times, each sample $s$ will have its own sets $\{r_k\}^{(s)}$ and $\{\tilde{r}_k\}^{(s)}$. We then consider the ensembles:
%	\begin{align}
%		\{\{r_k\}^{(s)}\} &= \{r_{(k,s)}\} \equiv \{r_l\} \equiv R \\
%		\{\{\tilde{r}_k\}^{(s)}\} &= \{\tilde{r}_{(k,s)}\} \equiv \{\tilde{r}_l\} \equiv \tilde{R} \label{eq_R_tilde}
%	\end{align}that is, the ensembles of all values of $r$ and the ensemble of all values of $\tilde{r}$ across all samples. We perform the calculations by exact diagonalization and compute 
%	 the probability distribution $P(r_l)$ of the gaps ratios for different values of $\lambda$. 
%	 The probability distribution $P(r_l)$ is computed by dividing the interval $\big[0,\max[R]\big]$ in bins of equal width $\Delta_{bin}$ and counting the number $n_k$ of elements of $R$ that fall within each bin $[k\Delta_{bin},(k+1)\Delta_{bin}]$. Here $k$ goes from $0$ to $N_{bin}-1$, where $N_{bin}=\frac{\max[R]}{\Delta_{bin}}$ is the total number of bins. The distribution $P(r_k)$ is thus given by $P(r_k) = \frac{n_k}{\abs{R}\Delta_{bin}}$, where $\abs{R}$ is the cardinality of $R$ (clearly, $\sum_k n_k = \abs{R}$ ). 

It has been shown \cite{Shaffer2014irreversibility,zhou2020single} that complex entanglement is associated with level repulsion in the entanglement spectrum and, specifically, with the emergence Wigner-Dyson (WD) statistics, namely the distribution of gap ratios \cite{atas2013distribution}
	\begin{align}
	    P_{\rm WD}(r) = \frac{1}{Z}\frac{(r+r^2)^\gamma}{(1+r+r^2)^{1+\frac{3}{2}\gamma}}~.
	\end{align}
This distribution of gap ratios is associated with random matrices belonging to the Gaussian Orthogonal Ensemble (GOE) (for which $\gamma = 1$ and $Z = \frac{8}{27}$); as well as the Gaussian Unitary Ensemble (GUE) (for which $\gamma = 2$ and $Z = \frac{4\pi}{81\sqrt{3}}$) \cite{Mehta_book}. In our case of interest, the reduced density matrix is real-valued (as so are the wavefunction coefficients) so that the relevant ensemble is expected to be the GOE. On the other hand, ``simple" entanglement is associated with departure from the Wigner-Dyson statistics, and the emergence of alternative statistics not showing level repulsion -- the simplest case being Poisson's statistics, associated with random eigenvalues, resulting in a gap-ratio distribution $P_{\rm Poisson}(r) = {(1+r)^{-2}}$~\cite{tang2021non-ergodic}.

\begin{figure}
		\centering
		\subfloat[]
  {\includegraphics[clip,width=\columnwidth]{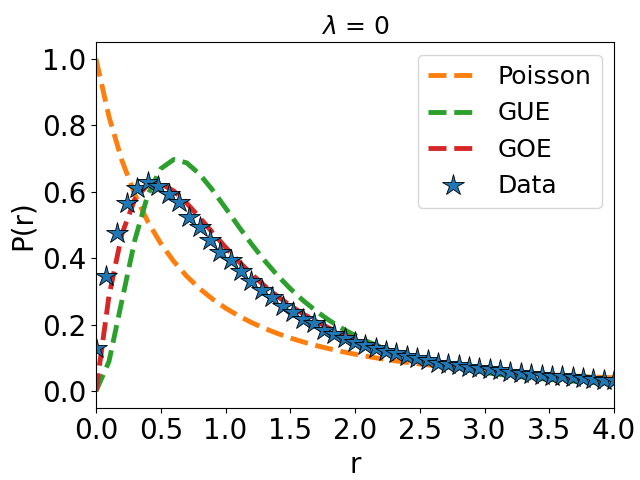}}
  
		\subfloat[]
  {\includegraphics[clip,width=\columnwidth]{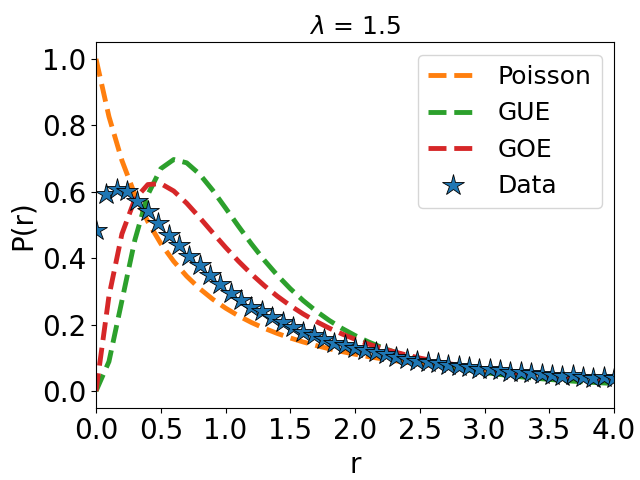}}
		\caption{
		Comparison between the distribution $P(r)$ for the gaps of the entanglement spectrum of $N=18$ qubits  and known statistical distributions (GOE, GUE and Poisson -- see text), for two different values of $\lambda$: (a) $\lambda = 0$; (b) $\lambda = 1.5$. 
		%In particular, the distributions considered are Wigner-Dyson distributions for the GOE  and the GUE, that show level repulsion ($P_{WD}(0) = 0$), and the Poisson distribution where level repulsion is absent ($P_{Poisson}(0) \neq 0$). 
%		Figure a) shows that, at $\lambda = 0$, the ESS (\emph{blue dots}) is in accordance with the GOE distribution (\emph{red line}) and manifests level repulsion, thus suggesting complex entanglement. 
%		On the other hand, figure b) shows that, at $\lambda = 1.5$, the ESS (\emph{blue dots}) does not possess level repulsion anymore, indicating non-complex entanglement, and a growing adherence to the Poisson distribution (\emph{orange line}). The distribution $P(r)$ is evaluated over $N_s = 500$ samples. 
		%The entanglement spectrum distribution $P(r)$ (blue dots) compared with different statistical distributions, in particular $P_{WD}(r)$ and $P_{Poisson}(r)$.  Figure a) shows the case for $\lambda = 0$, while figure b) shows the case for $\lambda = 1.5$. 
		}
		\label{fig_Pr}
	\end{figure}

In Fig.~\ref{fig_Pr} the gap-ratio distribution $P(r)$ (blue stars) is shown for two different values of $\lambda$ on the two sides of the entanglement transition. For $\lambda = 0 ~(< \lambda_E)$ we see that the gap distribution agrees well with the GOE Wigner-Dyson form, as expected from a complex-entanglement phase. On the other hand, for $\lambda=1.5 ~( > \lambda_E)$, the distribution strays away from the Wigner-Dyson form, although it is not complying either with Poisson statistics.  %???? HOW SHOULD ONE NOTE THIS: Note that the smaller $\lambda$ the more the system presents level repulsion.

\begin{figure}
		\centering
		\subfloat[]
  {\includegraphics[clip,width=\columnwidth]{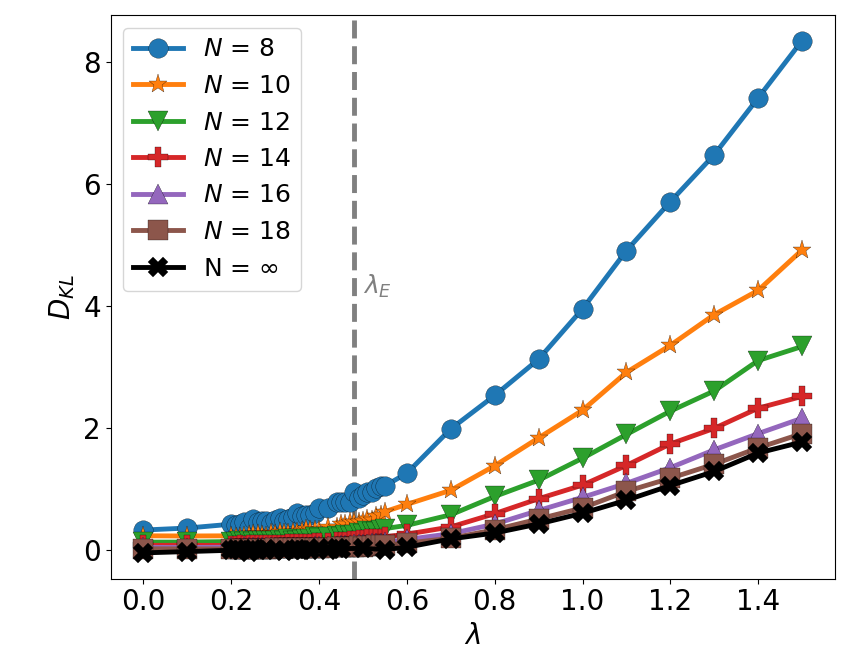}\label{fig_dkl_rt_a}}
  
		\subfloat[]
  {\includegraphics[clip,width=\columnwidth]{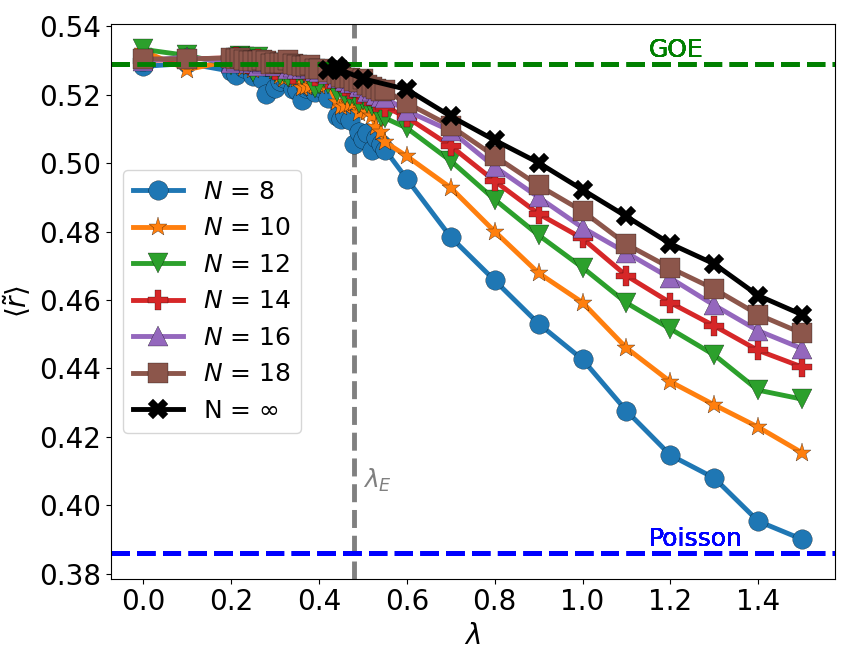}\label{fig_dkl_rt_b}}
		\caption{
		 Entanglement spectrum statistics of the RK-sign wavefunctions as a function of $\lambda$: (a) Kullback-Leibler divergence $D_{\rm KL}$ between the RK-sign distribution of gap ratios and the one stemming from Wigner-Dyson statistics. The dashed vertical line marks the critical value $\lambda_E$; (b) $\langle\tilde{r}\rangle$ as a function of $\lambda$, compared with the corresponding values for the GOE statistics ($\approx 0.53$) and the Poisson statistics ($\approx 0.39$). 
		% Again, for $\lambda < \lambda_c$, the curves converge, in the infinite-size limit, to the value of the GOE, while they stray away from it for larger values of $\lambda$. The black curves show the infinite $N$ behavior obtained from finite size scaling.
All results are obtained from a pool of $1000$ RK-sign wavefunctions.}
		\label{fig_dkl_rt}
	\end{figure}

To quantify the discordance between the gap-ratio distribution $P(r)$ of the RK-sign states and that resulting from Wigner-Dyson statistics, $P_{\rm WD}(r)$, we evaluate the {Kullback-Leibler divergence} between the two distributions
\begin{equation}
    D_{\rm KL} [P(r)||P_{\rm WD}(r)] = \sum_k P(r_k) \log_2[P(r_k)/P_{\rm WD}(r_k)].
\end{equation}
% Here, the values of $P_{WD}(r_k)$ are computed at the minimum of each bin. 
We show $D_{\rm KL}$ as a function of $\lambda$ in Fig.\ref{fig_dkl_rt_a} for different system sizes, along with an extrapolation to the thermodynamic limit. As one can see, for $\lambda \lesssim \lambda_E$, $D_{\rm KL}$ extrapolates to zero in the thermodynamic limit. Upon crossing the transition, $D_{\rm KL}$ becomes finite: this suggests that the entanglement transition of the RK-sign states is marked by a sharp shift in the entanglement spectrum statistics, namely it is accompanied by a sharp transition in entanglement complexity as well. 
% Then again we have a crossover that goes beyond the critical point, after which the $D_{\rm KL}$ assumes finite values and grows monotonically with $\lambda$. We see that the divergence $  D_{\rm KL}$  detects the two entanglement complexity phases. On the other hand, its behavior does not allow for a clear determination of the critical point $\lambda_c$. 

Another way to quantify the discordance of the entanglement spectrum statistics with the Wigner-Dyson statistics is to evaluate the average $\langle\tilde{r}\rangle$ defined in Eq.~\eqref{eq_R_tilde}. This average can be used as a direct indicator of the specific statistics followed by the entanglement spectrum, as it takes the value $\langle\tilde{r}\rangle \approx 0.53$ for the GOE; whereas \emph{e.g.} for Poisson statistics one has the value $\langle\tilde{r}\rangle\approx0.39$ \cite{huse2007localization}. $\langle\tilde{r}\rangle$  is plotted as a function of $\lambda$ in Fig.~\ref{fig_dkl_rt_b}, where one can see that %{\color{blue} 
for a certain $\lambda_r$%}
the phase with $\lambda<\lambda_r$ is essentially consistent with the GOE statistics, while for $\lambda>\lambda_r$ there is a gradual departure from GOE behavior.
% although Poisson statistics is not realized close the transition -- it may be realized (if at all) only for large values of $\lambda$.
%{\color{blue} 
The exact value of $\lambda_r$ at which this departure happens cannot be estimated precisely, due to the residual noise in $\langle\tilde{r}\rangle$ close to the transition. Our calculations show that $\langle\tilde{r}\rangle(\lambda = 0.44) = 0.528 \pm 0.003$ and $\langle\tilde{r}\rangle(\lambda = 0.5) = 0.524 \pm 0.002$, allowing us to locate the transition point in this interval of $\lambda$ values. This result is compatible with the value of the entanglement transition $\lambda_E \approx 0.48$, and leads us to consider the two points as equivalent: $\lambda_r \simeq \lambda_E$.%}
 These metrics suggest therefore that the whole volume-law phase of the RK-sign states exhibits universal Wigner-Dyson statistics of the entanglement spectrum.

%As we have seen also for the Stabilizer 2-Renyi Entropy, the point at which the crossing happens is compatible with the critical point of the fidelity metric $\lambda_c$.

	%The Von Neumann Entropy belongs to the family of $\alpha$-Renyi entropies, and it corresponds to the particular case where $\alpha = 1$. This family of Renyi entropies for the entanglement spectrum has been studied in \cite{chen2015mbl} where the critical behavior of these entropies identifies the transition to a Many-Body Localization phase. However, it is also shown that the critical point depends on both the value of $\alpha$ and the ratio $\frac{N_a}{N}$. Since the MBL analysis has been performed mostly with $\alpha \geq 2$ and $\frac{N_a}{N} = \frac{1}{3}$, the critical points shown in the reference differ from the one identified in this work.

\subsection{Ensemble fluctuations of the entanglement entropy}	
		
	Refs.~\cite{Leone2021quantum,oliviero2021transitions,true2022transitions} have shown that another signature of entanglement complexity in an ensemble of wavefunctions is offered by ensemble entanglement-entropy fluctuations. Following the analysis of the above-cited references, we define the variance of sample-to-sample fluctuations of the half-system entanglement entropy 
	\begin{equation}
		{\rm Var}(S_{N/2}) \equiv \overline{S^2_{N/2}(\psi)} - { \overline{S_{N/2}(\psi)} }~^2 ,
	\end{equation}
	where again the symbol $\overline{(...)}$ represents the average over a number $N_s$ of samples (here, $N_s = 1000$).

\begin{figure}
		\centering
		\includegraphics[clip,width=\columnwidth]{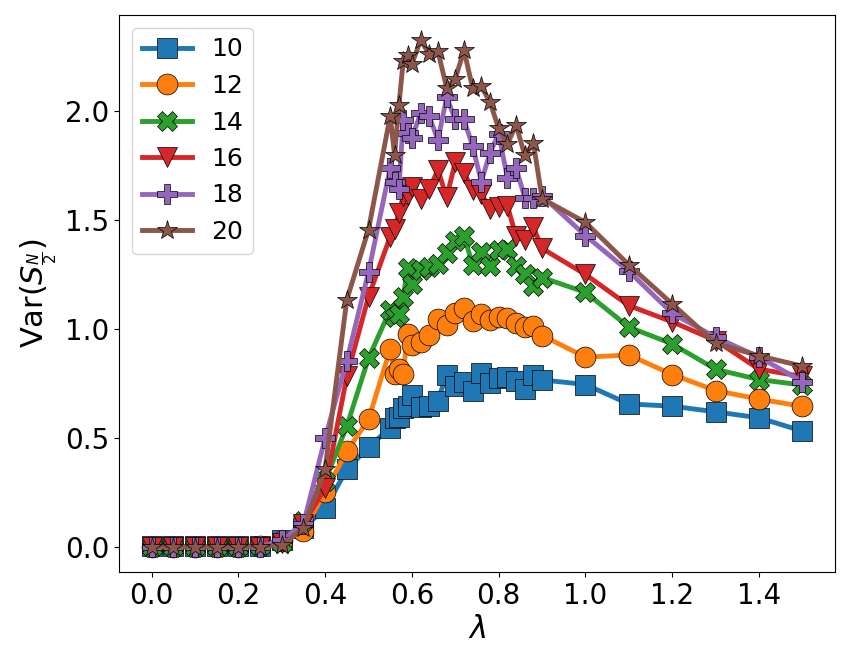}
		\caption{  
		Entanglement-entropy fluctuations of the half-bipartition of the system,  as functions of $\lambda$, for different system sizes. The averages computed in the derivation of ${\rm Var}(S_{N/2}) $ are obtained from a pool of $1000$ samples.
		}
		\label{fig_exp}
	\end{figure}

Fig.~\ref{fig_exp} shows the variance of the half-system entanglement entropy as a function of $\lambda$; a peak appears for intermediate $\lambda$ getting sharper with increasing system size, and moving to lower values of $\lambda$. While a scaling analysis of the peak position is rendered complicated by the noisy nature of the variance extracted from limited statistics, it appears rather plausible that the peak position drifts towards the entanglement transition at $\lambda_E$. Moreover the peak clearly separates rather sharply two scaling regimes: one of very slow scaling of the variance upon increasing the system size at small $\lambda$; and another one of much faster scaling at large $\lambda$.
	
\begin{figure}
		\centering
		\subfloat[]{\includegraphics[clip,width=\columnwidth]{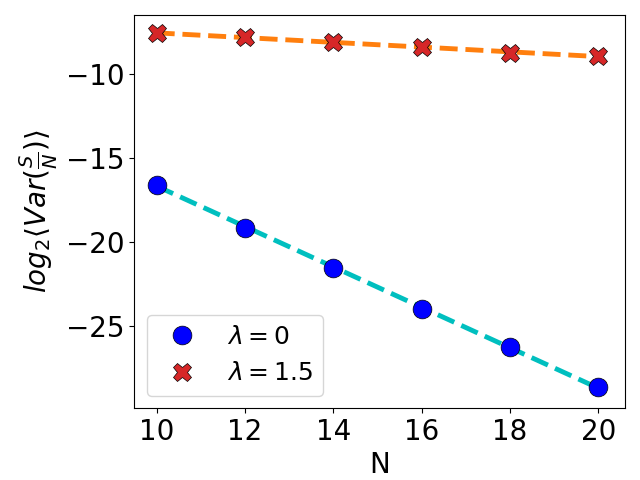}\label{fig_fluctuations_a}}
  
		\subfloat[]{\includegraphics[clip,width=\columnwidth]{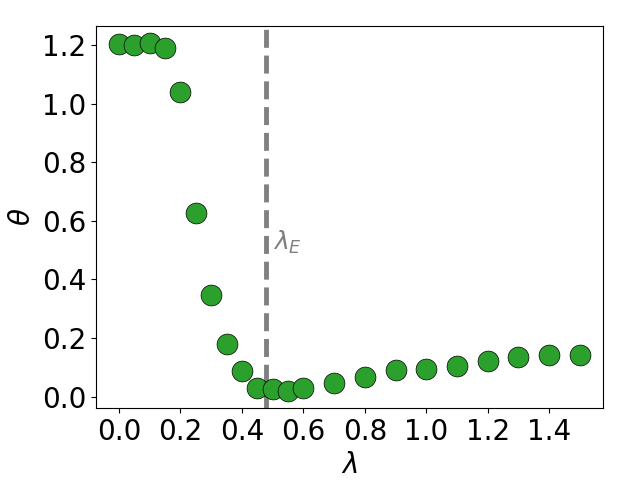}\label{fig_fluctuations_b}}
		\caption{
			(a)  Scaling of the entanglement entropy ensemble fluctuations with system size, fitted to the scaling Ansatz $\log_2 [{\rm Var}(S_{N/2}) / N] = - \theta N + {\rm const}$, for two values of $\lambda$. The dashed lines represent the best fit lines for each choice of $\lambda$: if  $\lambda = 0$, one finds $\theta = 1.205 \pm 0.009$; whereas if $\lambda = 1.5$, one finds $\theta = 0.140 \pm 0.006$. (b) Exponent $\theta$ for the scaling of entanglement-entropy fluctuations as a function of $\lambda$. 
		% The plot shows again a complex behavior, associated with a rapid scaling of the fluctuations with the system size, when $\lambda$ is small; however as $\lambda$ becomes larger the system transitions to a slower scaling, that is larger fluctuations, indicating simpler entanglement.
		}
		\label{fig_fluctuations}
	\end{figure}

 The scaling Ansatz capturing the size-dependence of the variance across the entire parameter range reads as
	 \begin{equation}
	 {\rm Var}(S_{N/2} / N) \propto d^{-\theta}
	 \end{equation}
	where $d \equiv 2^N$ is the Hilbert-space dimension.  The exponent $\theta$ can then be obtained from a linear fit of $\log_2[{\rm Var}(S_{N/2}/N)]$ vs. $N$ -- examples of such fits are shown in Fig.~\ref{fig_fluctuations_a} for two values of $\lambda$, providing widely different values of $\theta$ between the $\lambda < \lambda_E$ regime and the $\lambda > \lambda_E$ regime. 	
 Fig.~\ref{fig_fluctuations_b} puts together the values of the $\theta$ exponent obtained for different values of $\lambda$. In particular, two asymptotic regimes emerge: one for $\lambda\lesssim 0.2$ with $\theta \approx 1.2$, and one for $\lambda \gtrsim 1.5$ with $\theta \approx 0.14$. The first regime is in fact rather remarkable, as it is exhibits a \emph{universal} scaling of entanglement entropy fluctuations consistent with that of the output states of random quantum circuits, for which $\theta\sim 1.25$ \cite{true2022transitions}. 
These two regimes are separated by a sharp drop in $\theta$ which marks the transition for $\lambda \approx \lambda_E$, accompanied by the slowest scaling. This behavior is clearly reflected in the appearance of the peak in Fig.~\ref{fig_exp}.

%The study of random quantum circuits has shown that universal fluctuations for entanglement entropy are given by $\theta\sim 1.25$\cite{true2022transitions}, so our result shows that indeed the small $\lambda$ phase features entanglement fluctuations that are close to the universal ones. 

	%This is consistent with the ... {\bf FEATURE IN THE ENTANGLEMENT FLUCTUATIONS AS A FUNCTION OF $\lambda$ -- A PEAK EMERGING FOR LARGE $N$?}  
%	\begin{equation}
%		\log_2\frac{\langle Var_\psi\rangle}{N} \simeq -\theta N + \delta.
%	\end{equation}

%	In figure \ref{fig_exp} we plot the values that $\theta$ assumes for different $\lambda$, showing again the existence of two different behaviors: In the complex entangled phase where $\lambda<\lambda_c\simeq 0.2$ the fluctuations have a universal scaling, while as $\lambda>\lambda_c$  the scaling of the fluctuations  becomes non-universal.
%	In figure \ref{fig_fluctuations} we show the best fit lines obtained when one considers different values of $N$, for two different values of $\lambda$. The results show that, if $\lambda = 0$ then $\theta \simeq 1.205 \pm 0.009$, while if $\lambda = 1.5$ then $\theta \simeq 0.141  \pm 0.006$. This reveals, in analogy with the temporal fluctuations studied in \cite{true2022transitions}, a universal and non-universal scaling of the entropy fluctuations, respectively. 
	
\subsection{Stabilizer R\'enyi Entropy}

As we have seen in the previous sections, the entanglement transition of the RK-sign states -- from volume-law scaling to subextensive scaling --  is accompanied also by a transition in entanglement complexity as revealed by the entanglement spectrum statistics. A similar phenomenology has been observed for the high-energy excited states of the disordered XXZ model \cite{yang2017entanglement}. 
Nonetheless, it is important to remark that volume-law scaling for entanglement is not necessarily connected to a complex pattern of entanglement, as it has been shown in a series of works \cite{chamon2014emergent,Shaffer2014irreversibility,yang2015two-component,yang2017entanglement,zhou2020single,Leone2021quantum}. For instance, the output of a random quantum circuit made of Clifford gates features volume-law scaling for entanglement, but it exhibits Poisson entanglement spectrum statistics and non-universal fluctuations for the entanglement entropy \cite{Leone2021quantum,true2022transitions} . % {\bf WHAT ARE UNIVERSAL FLUCTUATIONS? THE RK-SIGN STATES DO NOT EXHIBIT A UNIVERSAL $\theta$ EITHER THROUGHOUT THE "COMPLEX" PHASE}. 
 The above example refers to states that have been produced with Clifford-group gates; therefore a direct way to tell apart such states with respect to the RK-sign ones is to probe the non-stabilizerness/magic of the latter. In order to do so,
%In some quantum many-body model, we mentioned that the ground state of local Hamiltonians can be high in resources beyond stabilizer states, the so-called magic, and low in entanglement\cite{oliviero2022magic}. In other words, magic and entanglement do not always accompany each other. One can have high entanglement with low magic (as in the output of Clifford circuits), or low entanglement and high magic, like in the ground state of the quantum Ising model.
%In order to obtain a complex, universal pattern of entanglement in the ESS, one needs to have both high magic and high entanglement.  The reason why the RK-sign model is interesting is that this is a quantum many-body model featuring a phase transition for both entanglement and magic at the same time, and therefore, we argue, a transition in entaglement complexity. 
%The resource beyond the stabilizer formalism, i.e., magic,  is  regarded as the critical resource that is responsible for quantum advantage\cite{bravyi2005universal} and is described in the framework of resource theory\cite{Veitch2014resource,howard2017application}. Recently, it has been shown that this quantity can be accessed more directly by 
we consider the entropy associated with the probability distribution over the Pauli decomposition of a state -- the so-called stabilizer R\'enyi entropy \cite{leone2022stabilizer}, which quantifies explicitly non-stabilizerness. 
%This way of computing non-stabilizer resources is particularly convenient for quantum many-body systems, where the interplay between magic and emergent order or complexity is starting to be considered\cite{liu2020many, Sarkar2020characterization,White2021cft, haug2022quantum}. 

\begin{figure}
		\centering
		\subfloat[]{\includegraphics[clip,width=\columnwidth]{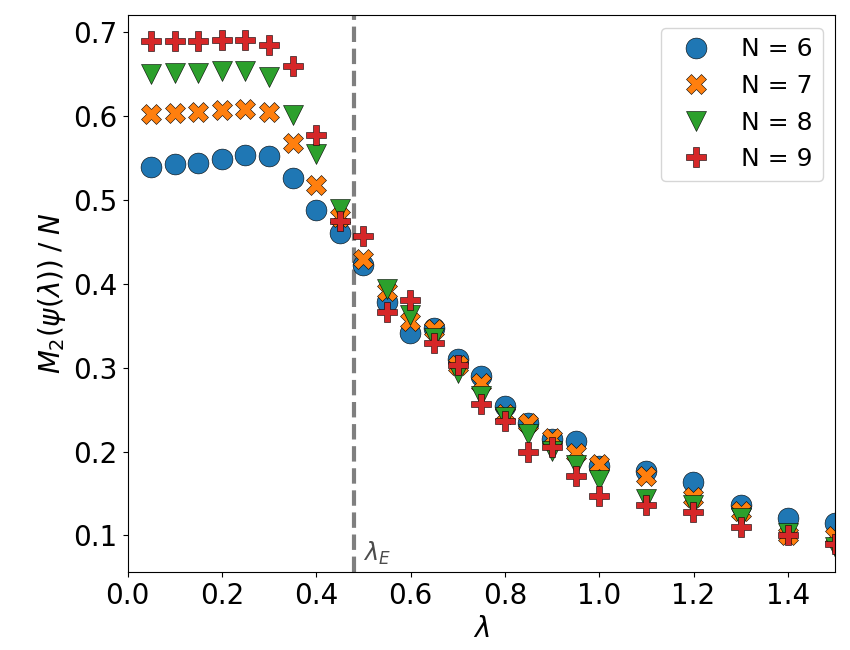}\label{fig_magic}}
  
		\subfloat[]{\includegraphics[clip,width=\columnwidth]{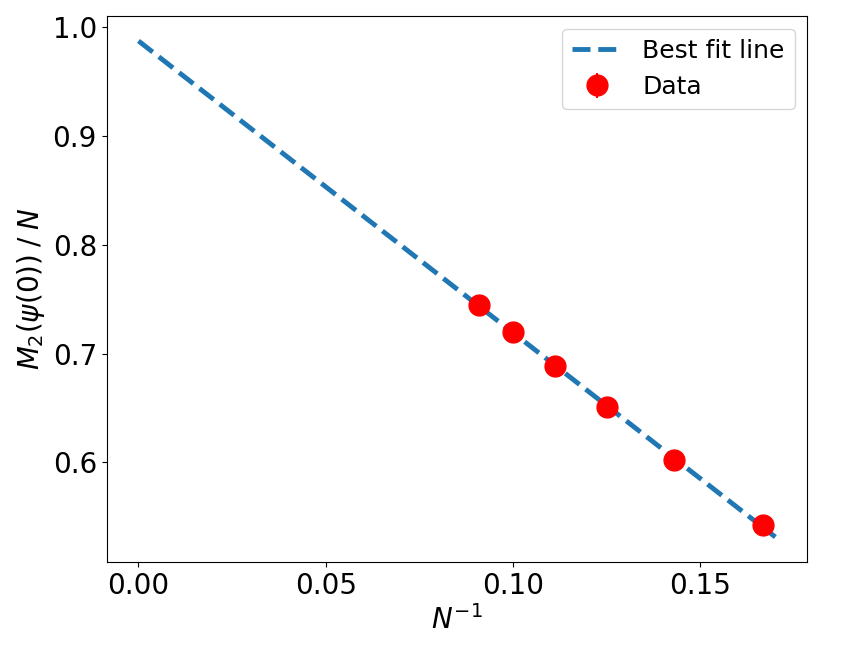}\label{fig_magic_scaling}}
		%\subfloat[]{\includegraphics[clip,width=0.5\columnwidth]{images/magic/M_vs_N_2.png}\label{fig_magic_N}}
		\caption{
		(a) Stabilizer Renyi entropy $M_2$ per qubit as a function of $\lambda$, for different system size; each point is computed averaging over $N_s = 400$ samples. (b) Scaling of Stabilizer Renyi entropy per qubit at $\lambda = 0$. The system size $N$  goes from $6$ to $11$. The number of realizations $N_s$ is $25$ for $N = 11$, $400$ otherwise. $R^2 = 0.9996$~\cite{virtanen2020scipy} indicates a good linear fit. The intercept of the best-fit line represents the value of $M_2(0)$ in the thermodynamic limit: $M_2(0) \simeq 0.99 N$.
		%(b) same quantity as in (a), but plotted this time as a function of size for a set of $\lambda$ values. In both panels each point is computed averaging over $N_s = 400$ samples. 
		}
		\label{fig_magic_all}
	\end{figure}
	
The stabilizer 2-R\'enyi entropy is defined as follows: consider the set $\mathcal{P}$ of all operators obtained from the tensor product of $N$ single-qubit Pauli gates -- that is $\mathcal{P} = \{\mathbb{I},X,Y,Z\}^{\otimes N}$. If $P_n$ ($n=1,\dots,4^N$) are the elements of $\mathcal{P}$, then the stabilizer 2-R\'enyi entropy $M_2$ of any pure state $\ket{\psi}$ is given by:
	\begin{equation}
	    M_2(\ket{\psi}) = -\log_2\sum^{4^N}_{n=1}\left(\frac{\bra{\psi}P_n \ket{\psi}^2}{2^N}\right)^2 - N.
	\end{equation}
	Note that the computation of this quantity requires to evaluate $4^N$ expectation values of $2^N \times 2^N$ operators, and that, for each value of $\lambda$, the random nature of the RK-sign states requires that we also average the result over several ($N_s$) samples. This means that $M_2(\lambda)$ requires significantly longer computation times with respect to other quantities that have been considered earlier in this work. It is therefore computationally expensive to reach the same system sizes as those considered in the previous sections, and to provide a proper finite-size scaling analysis. Nonetheless, already the accessible sizes we could consider ($N = 6,7,8,9$), provide a suggestive picture of the behavior of non-stabilizerness in the RK-sign states, as shown in Fig.~\ref{fig_magic}. There the curves of $M_2/N$ appear to exhibit a super-extensive scaling for sufficiently low $\lambda$ -- which, similarly to the entanglement entropy, must necessarily turn into extensive scaling at larger $N$, as one can prove that $M_2 \leq N$ for $N$ qubits \cite{leone2022stabilizer}. 
	In fact, as shown in %Appendix \ref{ap_magic}, 
	Fig.~\ref{fig_magic_scaling}, for $\lambda = 0$ we can shown that $M_2/N$ extrapolates to a value close to $0.99$, nearly saturating its upper bound.  We argue that this scaling behavior should persist over the entire regime $\lambda\lesssim 0.2$, given that the stabilizer R\'enyi entropy is essentially independent of $\lambda$ over this $\lambda$ range for all the system sizes we considered. 
		 
Increasing the value of $\lambda$, the  $M_2/N$ curves are found to  drop to smaller values, with a maximum $\lambda$-derivative (in absolute value) corresponding roughly to $\lambda_E$. Nonetheless, even upon entering the sub-volume-law phase, the scaling of the stabilizer R\'enyi entropy appears to maintain a volume law. The coexistence between an extensive  stabilizer R\'enyi entropy and subextensive entanglement entropy is not surprising, and it has been observed in the ground state of the quantum Ising chain \cite{oliviero2022magic}.

	\begin{figure}
		\centering
		\subfloat[]{\includegraphics[clip,width=\columnwidth]{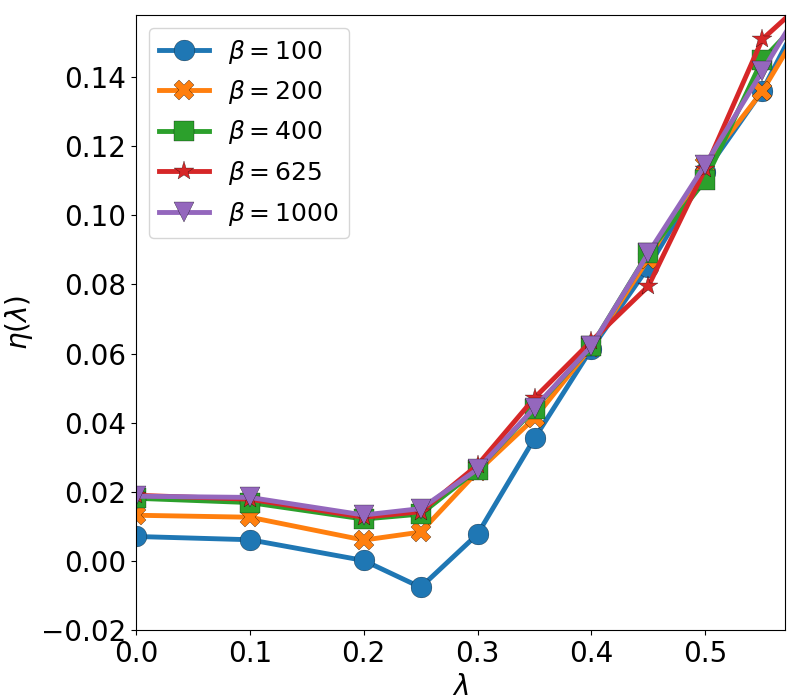}\label{fig_sched_a}}
  
		\subfloat[]{\includegraphics[clip,width=\columnwidth]{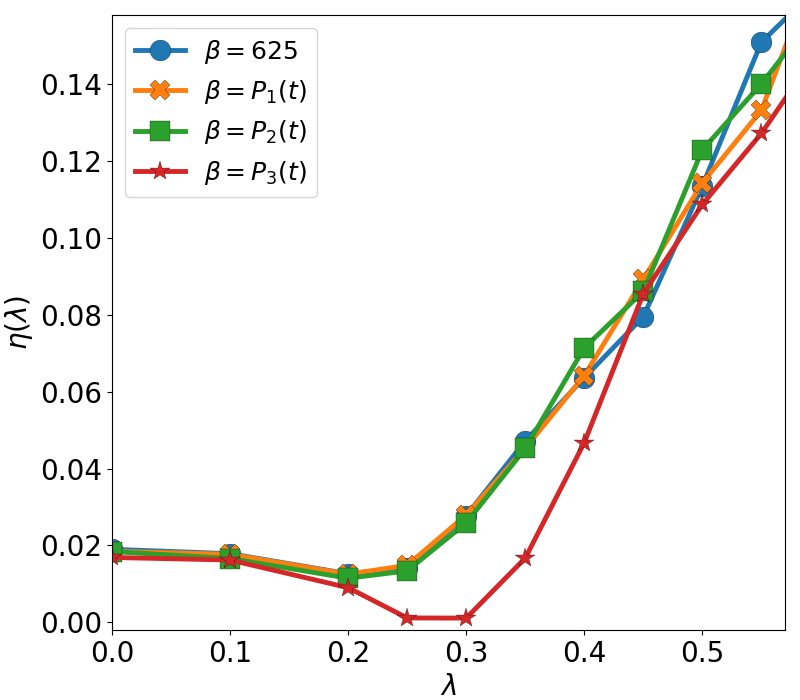}\label{fig_sched_b}}
		\caption{
		Efficiency $\eta$ of the disentangling algorithm applied to RK-sign states with $N=12$ qubits. (a) Constant-$\beta$ schedules, with different $\beta$ values. (b) Variable-$\beta$ schedules: blue (\emph{const.}): $\beta(t) = \beta_0$; orange: $\beta(t) = P_1(t) = \beta_0 (t+1)^{0.1}$; green: $\beta(t) = P_2(t) =  \beta_0 + 7.1\times 10^{-6} t^2$; red: $\beta(t) = P_3(t) =\beta_0 + 1.2\times 10^{-5} t^2$ . In both panels, the algorithm has a total duration of $t_{\rm MAX} = 7200$ steps, and the averages are performed over a sample of size $N_s = 100$.
		}
		\label{fig_sched}
	\end{figure}

\subsection{Disentangling hardness}

A final, more heuristic probe of the complexity of entanglement featured by the RK-sign states is offered by the efficiency of disentanglement using an entanglement annealing algorithm. Within this algorithm, the RK-sign states are evolved with a Clifford circuit stochastically built as follows. The random extraction of a single qubit gate (Hadamard gate, S gate) or a two-qubit (CNOT) gate is put to a Metropolis-algorithm test, namely if the gate leads to a decrease in entanglement entropy across the average over all the half-system bipartitions,  then the gate is retained. If instead the entanglement entropy increases by a $\Delta S$, then the gate is retained with a Metropolis-Hastings algorithm scheme, namely with a probability of acceptance given by $p = \exp(-\beta \Delta S)$. Here $\beta$ is a fictitious inverse temperature which is a parameter of the algorithm, and which can be a function of the annealing step (defining the so-called annealing schedule). The detailed scheme of the algorithm can be found in Ref.~\cite{Shaffer2014irreversibility,true2022transitions}.
 
 Several recent works have shown a link between entanglement complexity of a state and hardness to disentangle it with the above-cited algorithm. In Ref.~ \cite{chamon2014emergent,Shaffer2014irreversibility,zhou2020single,oliviero2021transitions,true2022transitions} on random quantum circuits, the presence of magic in the quantum states -- obtained for example by inserting $T-$gates in the random circuit -- is found to lead to output states of the circuit exhibiting Wigner-Dyson entanglement spectrum statistics. And, in turn, the same states are found to be hard to disentangle using the entanglement annealing algorithm -- namely the maximum amount of reduction of the von Neumann entropy in a given bipartition that can be obtained by the annealing algorithm is small compared to the initial entropy. A similar phenomenon accompanies the ETH-MBL transition in disordered systems, as shown in Ref.~\cite{yang2017entanglement}.

 Based on the above-cited works, we apply the disentangling-hardness diagnostics to the RK-sign states, expecting it to evolve significantly when moving across the entanglement transition. Introducing the initial half-system entanglement entropy ($S_i$) and the final one ($S_f$) after the disentangling procedure, we define the efficiency of the algorithm via the relative disentangling performance:
	\begin{equation}
	\eta(\lambda)= 1-\overline{S_f/S_i}.
	\end{equation}
	where the symbol $\overline{(...)}$ indicates again the average over a number $N_s$ of extractions of the initial state and subsequent realization of the annealing algorithm.
	In the following we take $N_s=200$, at which we observe a convergence of the averaged disentangling performance with the size of the sample.

	%%% Schedule evaluation
  The performance of the annealing algorithm depends in principle on 1) the annealing schedule, namely the variation of the effective inverse temperature $\beta$ with the number of annealing steps; and 2) the total number of annealing steps $t_{\rm MAX}$, and its scaling with system size. 
 Our results indicate nonetheless that, in the $\lambda\lesssim 0.2$ regime, the efficiency of the algorithm remains very low, regardless of its parameters.

\begin{figure}
		\centering
		\includegraphics[clip,width=\columnwidth]{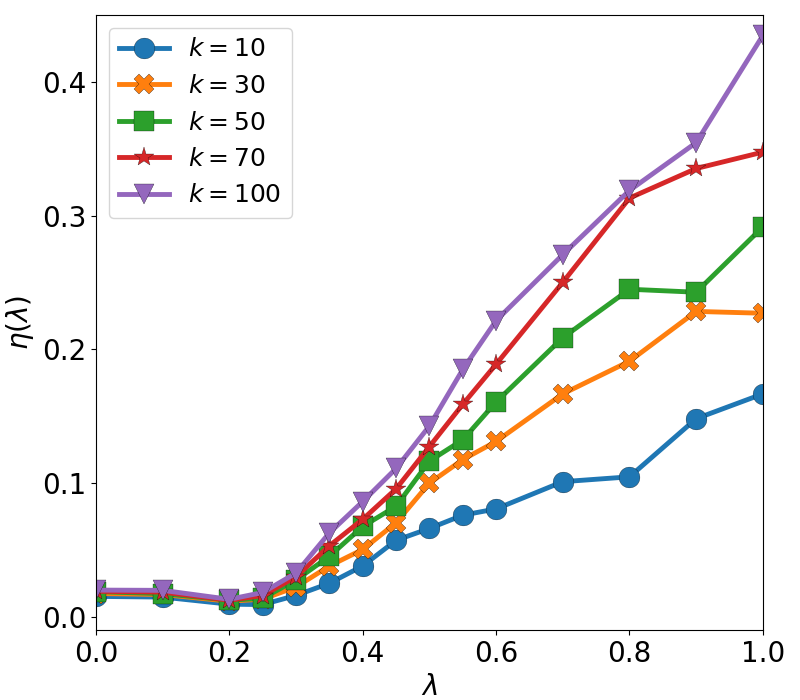}
		\caption{Dependence of the disentangling efficiency on the annealing schedule duration $t_{\rm MAX} = k N^2$. Here $N=12$ and $N_s=200$.}
		\label{fig_sched2}
	\end{figure}

Fig. \ref{fig_sched} shows $\eta(\lambda)$ for different annealing schedules, namely for various values of $\beta$ being held constant (Fig.~\ref{fig_sched_a}); and for different power-law annealing schedules in the form $\beta(t) = \beta_0 + \beta_1 t^\alpha$ in Fig.~\ref{fig_sched_b}. We observe that, regardless of the specific annealing schedule -- the length being fixed -- its performance remains very low in the small $\lambda$ regime on which the figures focus.
In particular the performance is minimal and nearly independent of $\lambda$ for $\lambda \lesssim 0.2$, which corresponds to a plateau regime for the entanglement entropy (Fig.~\ref{fig_ent_ent_a}),  the scaling of entanglement fluctuations (Fig.~\ref{fig_fluctuations}) and the stabilizer R\'enyi entropy (Fig.~\ref{fig_magic_all}). Given the small sensitivity on the annealing schedule, in the following we shall  concentrate on the simplest, constant-$\beta$ scheme;
in particular in Fig.~\ref{fig_sched_a} we observe that the disentangling efficiency stops increasing with $\beta$ for $\beta \gtrsim 400$; hence in the following we will consider $\beta = 400$, as this value falls in the regime that shows the saturation in efficiency. 

\begin{figure}
		\centering
		\includegraphics[clip,width=\columnwidth]{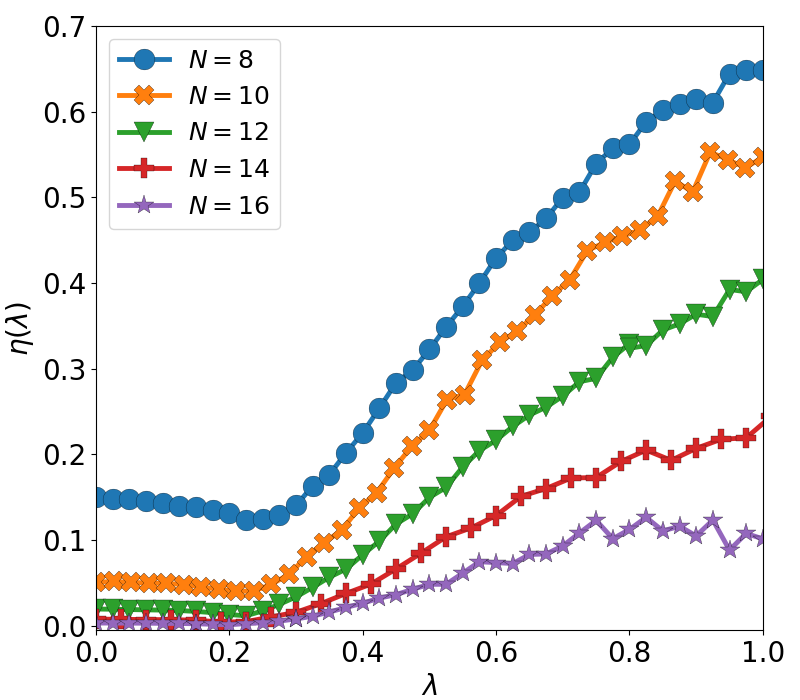}\label{fig_scaling_a}
		\caption{Dependence of the disentangling efficiency on system size for the constant-$\beta$ schedule. Here $N_s=200$ and $t_{\rm MAX}=100N^2$.}
		\label{fig_scaling}
	\end{figure}

As for the dependence on the overall length of the annealing schedule, following
	\cite{chamon2014emergent,Shaffer2014irreversibility,true2022transitions} we take  $t_{\rm MAX}(N)$ to scale as $t_{\rm MAX} = k N^2$, where $k$ is constant. In Fig.~\ref{fig_sched2} we plot $\eta(\lambda)$ for different values of $k$, observing again that for $\lambda \lesssim 0.2$ there is nearly no dependence of the disentangling efficiency on the length of the annealing protocol; while for larger values of $\lambda$ the efficiency goes up with $k$ -- marking a regime of partial disentanglement --  but the $k$-dependence becomes weak for $k \approx 100$.  A final remark concerns  the size dependence of the disentangling efficiency: Fig.~\ref{fig_scaling} shows that, when the annealing duration scales in the same way with $N^2$, the disentangling efficiency strongly decreases with increasing system size; in particular, for $\lambda \lesssim 0.2$, the efficiency is found to rapidly scale to zero upon increasing $N$. 
		
In conclusion, we observe that the disentangling efficiency is a good indicator of the higher complexity of entanglement of the RK-sign states for small $\lambda$ compared to those at large $\lambda$.  Neither the fidelity-metric transition at $\lambda_c$ nor the entanglement transition at $\lambda_E$ seem to affect the $\lambda$ dependence of the annealing efficiency. Yet the overall behavior  reveals again a special regime for $\lambda \lesssim 0.2$ in which the annealing efficiency is very weakly dependent on any parameter of the annealing schedule (including its duration); and it rapidly falls to zero with system size. On the other hand, longer entanglement annealing schedules lead to a larger disentangling efficiency in the complementary $\lambda \gtrsim 0.2$ regime.

%{\color{blue}
\subsection{Super-universal regime}
\begin{figure}[H]
		\centering
		\includegraphics[clip,width=\columnwidth]{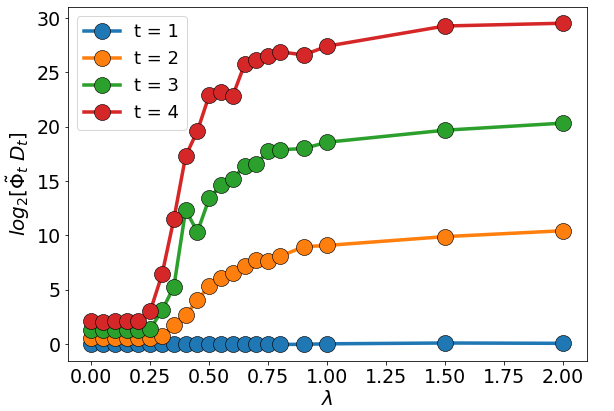}
		\caption{Logarithm of ${\tilde{\Phi}_t(\{\psi_i(\lambda)\})}{D_t}$ as a function of $\lambda$, for different values of the moment index ($t = 1, \dots, 4$). Each data-point corresponds to an ensemble of $K = 1000$ realizations of RK-sign states of $N = 12$ qubits.}
        \label{fig_FP_ratio}
\end{figure}

{ As seen in the previous sections, for $\lambda \lesssim 0.2$ we observe a super-universal regime, namely a regime in which the ensemble of states belonging to the RK-sign family exhibits an average entanglement, entanglement fluctuations and magic which scale in a way very similar to that of the ensemble of random states in Hilbert space.
Such states can be thought of as generated from a same reference state (say all $N$ spins up) which is acted upon by a unitary which is random in the sense of the Haar measure. Similarly one can view the RK-sign states as generated via random unitaries distributed according to a different measure than the Haar measure (hereafter called RK-sign measure). In this section we attempt at a more detailed comparison of the RK-sign measure with the Haar measure, aiming at better characterizing the super-universal regime and contrasting it with respect to the regimes at larger $\lambda$.
%The super-universal regime describes ensembles of states $\{|\psi(\lambda)\rangle\}$ that are random enough in the Haar sense from the entanglement point of view. In other words, the spectrum made of the eigenvalues of the reduced density matrix should be as close as possible as that of a random matrix. 

First of all, we can understand the similarity between the statistical properties of the RK-sign states at small $\lambda$ and the ones of random states }  by noticing that, in the limit $\lambda \rightarrow 0$, $|\psi(\lambda)\rangle$ is a superposition of all possible basis configurations with equal amplitudes and random signs. The amplitudes of the wave-function coefficients are fixed and the phase of the coefficients is extracted out of a bimodal distribution (with values 0 or $\pi$) instead of a fully flat one (on $[0,2\pi]$). Yet this appears to be sufficient in order to reproduce { the above-cited scaling of  entanglement and magic properties of fully random states.} In particular, this observation strongly suggests that the randomization of amplitudes is  irrelevant for entanglement complexity and that its defining aspect  is phase randomization.
 
 % A related question is whether the ensemble of pure states $\{|\psi_i(\lambda)\rangle\}_{i=1}^K$ (that is, the global states as opposed as the reduced density operators) is Haar-random as a function of $\lambda$. Of course, for large $\lambda$, such states are not close to random states as one obtains a superposition that is quite peaked around the states of minimum energy. As a consequence, one obtains area law for entanglement which we know not being typical of random states. However, in the $\lambda<\lambda_E$ regime one has volume law, which is typical of random states, and especially, universal ESS. The super-universal regime is detected by yet more fine tuned properties of entanglement such as the entanglement fluctuations, the disentangling hardness, and the stabilizer entropy behavior. 
 
 { Moving beyond these qualitative observations, we compare the Haar-measure ensemble with the RK-sign ensemble by evaluating whether the latter makes up for an approximate $t-$design \cite{zhu2016clifford}, namely, whether the RK-sign measure is capable of reproducing the statistics of the Haar measure of up to the $t$-th moment}. As it has been shown in Ref.~\cite{Leone2021quantum}, an important threshold is that of $t=4$, which sharply discriminates the Clifford group from the Haar measure and it is at the onset of quantum chaos. The ensemble $\{|\psi_i(\lambda)\rangle\}_{i=1}^K$ { -- with $K$ the number of samples -- } is a (projective) $t-$design if and only if its $t-$th frame potential, defined as \cite{zhu2016clifford}
 \begin{equation}\label{frame_potential}
    \Phi_t(\{\psi_i(\lambda)\}_{i=1}^K) = \frac{1}{K^2}\sum_{j,k}\abs{\braket{\psi_j|\psi_k}}^{2t}.
\end{equation}
 attains the value 
 \begin{equation}
 \Phi_t(\{\psi_i(\lambda)\}_{i=1}^K) = D_t^{-1} \equiv \binom{d+t-1}{t}^{-1} 
 \end{equation}
 where $d=2^N$ is the dimension of the Hilbert space. 
 %Introducing a finite $\lambda$ has instead the effect of randomizing the amplitudes, with a probability distribution which acquires an increasingly fat tail towards small values. For moderate $\lambda$ values, this deformation of the wave-function is actually making the RK-sign states closer to random states, for which amplitudes are equally distributed over a finite interval. But when $\lambda$ becomes too large, the distribution of amplitudes, which was peaked around $1$ (up to normalization) at small $\lambda$, develops instead a peak around zero, and the statistics of RK-sign wave-functions starts departing from that of the random-state ensemble... {\bf CAN WE CHECK FOR WHICH $\lambda$ this happens? Maybe for $\lambda = 0.2?$}
%
%\subsection{RK-sign wavefunctions ensembles as $t$-designs}
%We consider a random ensemble $\{\psi_i(\lambda)\}$ of $K$ RK-sign states, and we want to determine if this set is a $t$-design. Following reference \cite{zhu2016clifford}, we need to compute its $t$th frame potential:
%\begin{equation}\label{frame_potential}
 %   \Phi(\{\psi_i(\lambda)\}) = \frac{1}{K^2}\sum_{j,k}\abs{\braket{\psi_j|\psi_k}}^{2t}.
%\end{equation}
In general, one has $\Phi_t(\{\psi_i(\lambda)\}_{i=1}^K) \geq {D_t}^{-1}$. { An ensemble with the property $D_t\Phi_t \simeq 1$ realizes} an approximate $t-$design. { On the contrary,} high values of the product $D_t\Phi_t$ denote low randomness of the ensemble of wave-functions.

The off-diagonal contributions to $\Phi_t$ can be written as $\tilde{\Phi}_t(\{\psi_i(\lambda)\}_{i=1}^K) = \Phi_t(\{\psi_i(\lambda)\}_{i=1}^K) - {K}^{-1}$. Numerical evaluation shows that  the number $\tilde{\Phi}$ converges rapidly with  $K$, which means that it also well represents the limit $K\rightarrow\infty$ for the frame potential $\Phi_t$.

%and the equality holds iff $\{\psi_i(\lambda)\}$ is a $t$-design. However, note that, in equation \ref{frame_potential}, the terms $\abs{\braket{\psi_j|\psi_k}}^{2t}$ are all $1$ if $j=k$, no matter what is the ensemble that is considered, and account for a term $\frac{1}{K}$ in the overall summation. This means that, if $K$ is not large enough we will always have $\Phi(\{\psi_i(\lambda)\}) \approx \frac{1}{K}$. define
%\begin{equation}    
%\tilde{\Phi}(\{\psi_i(\lambda)\}) = \Phi(\{\psi_i(\lambda)\}) - \frac{1}{K},
%\end{equation} 
%and we compare this with $1 / D_t$. 
In figure \ref{fig_FP_ratio} we plot the logarithm of  ${\tilde{\Phi}_t(\{\psi_i(\lambda)\}_{i=1}^K)}{D_t}$ for a sample $\{\psi_i(\lambda)\}_{i=1}^K$ of $K = 1000$ realizations of RK-sign states of $N = 12$ qubits, as a function of $\lambda$, and for different values of the moment $t$. As we can see, the super-universal region shows a constant behavior of the $t$-frame potential, which has the same order of magnitude of  the value for a $4-$design. In this sense this is an approximate $4-$design. In comparison, as one crosses over away from the super-universal region, the value of $D_t\Phi_t$ rapidly increases to very high values, as one can see from a logarithmic scale, see Fig.\ref{fig_FP_ratio}. This signals a marked  departure from a projective $t-$design, the more pronounced the higher the moment $t$.

%As we can see, the super-universal region shows a constant behavior of the $t$-frame potential, {\color{red} and it is compatible with an approximate $t-$design with $t$ up to 4 -- CAN WE SAY THIS EVEN IF THE PARAMETER $D_t\Phi_t$ TAKES VALUES $\sim 8$?} . As we cross over away from the super-universal region, we have a marked {\color{red} departure} from a projective design {\color{red} (the more pronounced the higher the moment)}
%, signaling that here the global states are far from being Haar-random. 
We observe therefore that the the super-universal regime features high (albeit imperfect) randomness of the ensemble of wavefunctions, together with a nearly perfect independence of such randomness with respect to the parameter $\lambda$.
%: it is a randomness-robust region. 
%}

	\section{Discussion}
      \label{s.discussion}

   In summary, our analysis of the entanglement features of the RK-sign wavefunctions shows that this family of states possesses an entanglement transition decoupled from that of the fidelity metric; and separating a volume-law phase, characterized by universal Wigner-Dyson statistics for the entanglement spectrum, from a sub-volume-law phase with non-universal entanglement spectrum statistics. Within the volume-law phase a super-universal regime appears characterized by universal entanglement entropy fluctuations, stabilizer R\'enyi entropy scaling towards its theoretical maximum, and close-to-null efficiency of a disentangling algorithm. These results show that the RK-sign states showcase a very rich palette of different levels of entanglement complexity, essentially encompassing all the regimes observed previously piecewise in various model systems -- from the output states of random quantum circuits, to ground and excited states of many-body Hamiltonians. 
   
   The phenomenology unveiled here in a one-parameter family of model quantum states may be expected to appear in the entanglement features of excited Hamiltonian eigenstates -- a strong candidate being the excited states of disordered Hamiltonians featuring a many-body localization transition. A systematic study of entanglement complexity metrics for such states (and in particular of the stabilizer R\'enyi entropy), going beyond the analysis offered by Ref.~\cite{yang2017entanglement}  appears therefore as a rather appealing perspective. Nonetheless it is important to underline the significant computational overhead required to extract the stabilizer R\'enyi entropy of quantum states; on the technical side, more efficient approaches may be envisioned to calculate it, based \emph{e.g.} on a stochastic sampling of the Pauli decompositions of a state.

\section{Acknowledgments}
SP and AH acknowledge support from NSF award number 2014000, as well as the hospitality of the ``Laboratoire de Physique" of the ENS of Lyon, where part of this work was performed. The work of SP was supported in part by College of Science and Mathematics Dean’s Doctoral Research Fellowship through fellowship support from Oracle, project ID R20000000025727. SP would also like to thank Lorenzo Leone and Salvatore Francesco Emanuele Oliviero for providing helpful discussions in the production of this work. TR is supported by ANR (`EELS' project) and QuantERA (`MAQS' project). AH acknowledges financial support from PNRR MUR project PE0000023-NQSTI.
	%\newpage
		
	%\bibliography{biblio.bib}

\begin{thebibliography}{10}
\providecommand{\url}[1]{\texttt{#1}}
\providecommand{\urlprefix}{URL }
\expandafter\ifx\csname urlstyle\endcsname\relax
  \providecommand{\doi}[1]{doi:\discretionary{}{}{}#1}\else
  \providecommand{\doi}{doi:\discretionary{}{}{}\begingroup
  \urlstyle{rm}\Url}\fi
\providecommand{\eprint}[2][]{\url{#2}}

\bibitem{bell2004speakable}
J.~S. Bell and A.~Aspect,
\newblock \emph{Speakable and Unspeakable in Quantum Mechanics: Collected
  Papers on Quantum Philosophy},
\newblock Cambridge University Press, 2 edn.,
\newblock \doi{10.1017/CBO9780511815676} (2004).

\bibitem{Scaranibook}
V.~Scarani,
\newblock \emph{Bell nonlocality},
\newblock Oxford University Press (2019).

\bibitem{bennett1993teleporting}
C.~H. Bennett, G.~Brassard, C.~Cr\'epeau, R.~Jozsa, A.~Peres and W.~K.
  Wootters,
\newblock \emph{Teleporting an unknown quantum state via dual classical and
  einstein-podolsky-rosen channels},
\newblock Phys. Rev. Lett. \textbf{70}, 1895 (1993),
\newblock \doi{10.1103/PhysRevLett.70.1895}.

\bibitem{Nielsen}
M.~A. Nielsen and I.~L. Chuang,
\newblock \emph{Quantum information theory}, pp. 26--28,
\newblock Cambridge University Press,
\newblock \doi{10.1017/CBO9780511976667.016} (2010).

\bibitem{shor1994algorithms}
P.~Shor,
\newblock \emph{Algorithms for quantum computation: discrete logarithms and
  factoring},
\newblock In \emph{Proceedings 35th Annual Symposium on Foundations of Computer
  Science}, pp. 124--134,
\newblock \doi{10.1109/SFCS.1994.365700} (1994).

\bibitem{masanes2006all}
L.~Masanes,
\newblock \emph{All bipartite entangled states are useful for information
  processing},
\newblock Phys. Rev. Lett. \textbf{96}, 150501 (2006),
\newblock \doi{10.1103/PhysRevLett.96.150501}.

\bibitem{amico2008entanglement}
L.~Amico, R.~Fazio, A.~Osterloh and V.~Vedral,
\newblock \emph{Entanglement in many-body systems},
\newblock Rev. Mod. Phys. \textbf{80}, 517 (2008),
\newblock \doi{10.1103/RevModPhys.80.517}.

\bibitem{eisert2010colloquium}
J.~Eisert, M.~Cramer and M.~B. Plenio,
\newblock \emph{Colloquium: Area laws for the entanglement entropy},
\newblock Rev. Mod. Phys. \textbf{82}, 277 (2010),
\newblock \doi{10.1103/RevModPhys.82.277}.

\bibitem{hamma2012quantum}
A.~Hamma, S.~Santra and P.~Zanardi,
\newblock \emph{Quantum entanglement in random physical states},
\newblock Phys. Rev. Lett. \textbf{109}, 040502 (2012),
\newblock \doi{10.1103/PhysRevLett.109.040502}.

\bibitem{hamma2012ensembles}
A.~Hamma, S.~Santra and P.~Zanardi,
\newblock \emph{Ensembles of physical states and random quantum circuits on
  graphs},
\newblock Phys. Rev. A \textbf{86}, 052324 (2012),
\newblock \doi{10.1103/PhysRevA.86.052324}.

\bibitem{Osterloh2002scaling}
A.~Osterloh, L.~Amico, G.~Falci and R.~Fazio,
\newblock \emph{Scaling of entanglement close to a quantum phase transition},
\newblock Nature \textbf{416}(6881), 608 (2002),
\newblock \doi{10.1038/416608a}.

\bibitem{osborne2002entanglement}
T.~J. Osborne and M.~A. Nielsen,
\newblock \emph{Entanglement in a simple quantum phase transition},
\newblock Phys. Rev. A \textbf{66}, 032110 (2002),
\newblock \doi{10.1103/PhysRevA.66.032110}.

\bibitem{vidal2003entanglement}
G.~Vidal, J.~I. Latorre, E.~Rico and A.~Kitaev,
\newblock \emph{Entanglement in quantum critical phenomena},
\newblock Phys. Rev. Lett. \textbf{90}, 227902 (2003),
\newblock \doi{10.1103/PhysRevLett.90.227902}.

\bibitem{wu2004quantum}
L.-A. Wu, M.~S. Sarandy and D.~A. Lidar,
\newblock \emph{Quantum phase transitions and bipartite entanglement},
\newblock Phys. Rev. Lett. \textbf{93}, 250404 (2004),
\newblock \doi{10.1103/PhysRevLett.93.250404}.

\bibitem{roscilde2004studying}
T.~Roscilde, P.~Verrucchi, A.~Fubini, S.~Haas and V.~Tognetti,
\newblock \emph{Studying quantum spin systems through entanglement estimators},
\newblock Phys. Rev. Lett. \textbf{93}, 167203 (2004),
\newblock \doi{10.1103/PhysRevLett.93.167203}.

\bibitem{hofmann2014scaling}
M.~Hofmann, A.~Osterloh and O.~G\"uhne,
\newblock \emph{Scaling of genuine multiparticle entanglement close to a
  quantum phase transition},
\newblock Phys. Rev. B \textbf{89}, 134101 (2014),
\newblock \doi{10.1103/PhysRevB.89.134101}.

\bibitem{amico2009entanglementmagnetic}
L.~Amico and R.~Fazio,
\newblock \emph{Entanglement and magnetic order},
\newblock Journal of Physics A: Mathematical and Theoretical \textbf{42}(50),
  504001 (2009),
\newblock \doi{10.1088/1751-8113/42/50/504001}.

\bibitem{hamma2005ground}
A.~Hamma, R.~Ionicioiu and P.~Zanardi,
\newblock \emph{Ground state entanglement and geometric entropy in the kitaev
  model},
\newblock Physics Letters A \textbf{337}(1), 22 (2005),
\newblock \doi{https://doi.org/10.1016/j.physleta.2005.01.060}.

\bibitem{hamma2008entanglement}
A.~Hamma, W.~Zhang, S.~Haas and D.~A. Lidar,
\newblock \emph{Entanglement, fidelity, and topological entropy in a quantum
  phase transition to topological order},
\newblock Phys. Rev. B \textbf{77}, 155111 (2008),
\newblock \doi{10.1103/PhysRevB.77.155111}.

\bibitem{kitaev2006topological}
A.~Kitaev and J.~Preskill,
\newblock \emph{Topological entanglement entropy},
\newblock Phys. Rev. Lett. \textbf{96}, 110404 (2006),
\newblock \doi{10.1103/PhysRevLett.96.110404}.

\bibitem{Wen2007quantum}
X.-G. Wen,
\newblock \emph{Quantum Field Theory of Many-Body Systems: From the Origin of
  Sound to an Origin of Light and Electrons},
\newblock Oxford Graduate Texts. Oxford University Press, Oxford,
\newblock ISBN 9780199227259,
\newblock \doi{10.1093/acprof:oso/9780199227259.001.0001} (2007).

\bibitem{castelnuovo2008quantum}
C.~Castelnovo and C.~Chamon,
\newblock \emph{Quantum topological phase transition at the microscopic level},
\newblock Phys. Rev. B \textbf{77}, 054433 (2008),
\newblock \doi{10.1103/PhysRevB.77.054433}.

\bibitem{bardarson2012unbounded}
J.~H. Bardarson, F.~Pollmann and J.~E. Moore,
\newblock \emph{Unbounded growth of entanglement in models of many-body
  localization},
\newblock Phys. Rev. Lett. \textbf{109}, 017202 (2012),
\newblock \doi{10.1103/PhysRevLett.109.017202}.

\bibitem{vosk2013many-body}
R.~Vosk and E.~Altman,
\newblock \emph{Many-body localization in one dimension as a dynamical
  renormalization group fixed point},
\newblock Phys. Rev. Lett. \textbf{110}, 067204 (2013),
\newblock \doi{10.1103/PhysRevLett.110.067204}.

\bibitem{serbyn2013universal}
M.~Serbyn, Z.~Papi\ifmmode~\acute{c}\else \'{c}\fi{} and D.~A. Abanin,
\newblock \emph{Universal slow growth of entanglement in interacting strongly
  disordered systems},
\newblock Phys. Rev. Lett. \textbf{110}, 260601 (2013),
\newblock \doi{10.1103/PhysRevLett.110.260601}.

\bibitem{serbyn2013local}
M.~Serbyn, Z.~Papi\ifmmode~\acute{c}\else \'{c}\fi{} and D.~A. Abanin,
\newblock \emph{Local conservation laws and the structure of the many-body
  localized states},
\newblock Phys. Rev. Lett. \textbf{111}, 127201 (2013),
\newblock \doi{10.1103/PhysRevLett.111.127201}.

\bibitem{serbyn2014interferometric}
M.~Serbyn, M.~Knap, S.~Gopalakrishnan, Z.~Papi\ifmmode~\acute{c}\else
  \'{c}\fi{}, N.~Y. Yao, C.~R. Laumann, D.~A. Abanin, M.~D. Lukin and E.~A.
  Demler,
\newblock \emph{Interferometric probes of many-body localization},
\newblock Phys. Rev. Lett. \textbf{113}, 147204 (2014),
\newblock \doi{10.1103/PhysRevLett.113.147204}.

\bibitem{chen2015mbl}
X.~Chen, X.~Yu, G.~Y. Cho, B.~K. Clark and E.~Fradkin,
\newblock \emph{Many-body localization transition in rokhsar-kivelson-type wave
  functions},
\newblock Phys. Rev. B \textbf{92}, 214204 (2015),
\newblock \doi{10.1103/PhysRevB.92.214204}.

\bibitem{abanin2019colloquium}
D.~A. Abanin, E.~Altman, I.~Bloch and M.~Serbyn,
\newblock \emph{Colloquium: Many-body localization, thermalization, and
  entanglement},
\newblock Rev. Mod. Phys. \textbf{91}, 021001 (2019),
\newblock \doi{10.1103/RevModPhys.91.021001}.

\bibitem{szalay2012partial}
S.~Szalay and Z.~K\"ok\'enyesi,
\newblock \emph{Partial separability revisited: Necessary and sufficient
  criteria},
\newblock Phys. Rev. A \textbf{86}, 032341 (2012),
\newblock \doi{10.1103/PhysRevA.86.032341}.

\bibitem{walter2016multipartite}
M.~Walter, D.~Gross and J.~Eisert,
\newblock \emph{Multipartite Entanglement}, chap.~14, pp. 293--330,
\newblock John Wiley and Sons, Ltd,
\newblock ISBN 9783527805785,
\newblock \doi{https://doi.org/10.1002/9783527805785.ch14} (2016).

\bibitem{bengtsson2017multipartite}
I.~Bengtsson and K.~Życzkowski,
\newblock \emph{Multipartite entanglement}, p. 493–543,
\newblock Cambridge University Press, 2 edn.,
\newblock \doi{10.1017/9781139207010.018} (2017).

\bibitem{chamon2014emergent}
C.~Chamon, A.~Hamma and E.~R. Mucciolo,
\newblock \emph{Emergent irreversibility and entanglement spectrum statistics},
\newblock Physical Review Letters \textbf{112}, 240501 (2014),
\newblock \doi{10.1103/PhysRevLett.112.240501}.

\bibitem{Shaffer2014irreversibility}
D.~Shaffer, C.~Chamon, A.~Hamma and E.~R. Mucciolo,
\newblock \emph{Irreversibility and entanglement spectrum statistics in quantum
  circuits},
\newblock Journal of Statistical Mechanics: Theory and Experiment
  \textbf{2014}(12), P12007 (2014),
\newblock \doi{10.1088/1742-5468/2014/12/p12007}.

\bibitem{Roberts2017chaos}
D.~A. Roberts and B.~Yoshida,
\newblock \emph{Chaos and complexity by design},
\newblock Journal of High Energy Physics \textbf{2017}(4), 121 (2017),
\newblock \doi{10.1007/JHEP04(2017)121}.

\bibitem{boykin2000anew}
P.~Boykin, T.~Mor, M.~Pulver, V.~Roychowdhury and F.~Vatan,
\newblock \emph{A new universal and fault-tolerant quantum basis},
\newblock Information Processing Letters \textbf{75}(3), 101 (2000),
\newblock \doi{https://doi.org/10.1016/S0020-0190(00)00084-3}.

\bibitem{gottesman2009anintroduction}
D.~Gottesman,
\newblock \emph{An introduction to quantum error correction and fault-tolerant
  quantum computation},
\newblock Quantum information science and its contributions to mathematics, vol. 68 of Proc. Sympos. Appl. Math., pp. 13–58. Amer. Math. Soc., Providence, RI (2010)
\newblock \doi{10.1090/psapm/068/2762145}.

\bibitem{Gottesman:1998hu}
D.~{Gottesman},
\newblock \emph{{The Heisenberg Representation of Quantum Computers}},
\newblock arXiv (1998),
\newblock \href{http://arxiv.org/abs/quant-ph/9807006}{{
  [quant-ph/quant-ph/9807006]}}.

\bibitem{bravyi2005universal}
S.~Bravyi and A.~Kitaev,
\newblock \emph{Universal quantum computation with ideal clifford gates and
  noisy ancillas},
\newblock Phys. Rev. A \textbf{71}, 022316 (2005),
\newblock \doi{10.1103/PhysRevA.71.022316}.

\bibitem{Knill2005quantum}
E.~Knill,
\newblock \emph{Quantum computing with realistically noisy devices},
\newblock Nature \textbf{434}(7029), 39 (2005),
\newblock \doi{10.1038/nature03350}.

\bibitem{Veitch2014resource}
V.~Veitch, S.~A.~H. Mousavian, D.~Gottesman and J.~Emerson,
\newblock \emph{The resource theory of stabilizer quantum computation},
\newblock New Journal of Physics \textbf{16}(1), 013009 (2014),
\newblock \doi{10.1088/1367-2630/16/1/013009}.

\bibitem{howard2017application}
M.~Howard and E.~Campbell,
\newblock \emph{Application of a resource theory for magic states to
  fault-tolerant quantum computing},
\newblock Phys. Rev. Lett. \textbf{118}, 090501 (2017),
\newblock \doi{10.1103/PhysRevLett.118.090501}.

\bibitem{leone2022stabilizer}
L.~Leone, S.~F.~E. Oliviero and A.~Hamma,
\newblock \emph{Stabilizer r\'enyi entropy},
\newblock Phys. Rev. Lett. \textbf{128}, 050402 (2022),
\newblock \doi{10.1103/PhysRevLett.128.050402}.

\bibitem{Leone2021quantum}
L.~Leone, S.~F.~E. Oliviero, Y.~Zhou and A.~Hamma,
\newblock \emph{Quantum {C}haos is {Q}uantum},
\newblock {Quantum} \textbf{5}, 453 (2021),
\newblock \doi{10.22331/q-2021-05-04-453}.

\bibitem{true2022transitions}
S.~True and A.~Hamma,
\newblock \emph{Transitions in {E}ntanglement {C}omplexity in {R}andom
  {C}ircuits},
\newblock {Quantum} \textbf{6}, 818 (2022),
\newblock \doi{10.22331/q-2022-09-22-818}.

\bibitem{leone2021isospectral}
L.~Leone, S.~F.~E. Oliviero and A.~Hamma,
\newblock \emph{Isospectral twirling and quantum chaos},
\newblock Entropy \textbf{23}(8) (2021),
\newblock \doi{10.3390/e23081073}.

\bibitem{oliviero2021random}
S.~F.~E. Oliviero, L.~Leone, F.~Caravelli and A.~Hamma,
\newblock \emph{{Random matrix theory of the isospectral twirling}},
\newblock SciPost Phys. \textbf{10}, 076 (2021),
\newblock \doi{10.21468/SciPostPhys.10.3.076}.

\bibitem{leone2022magic}
L.~Leone, S.~F.~E. Oliviero and A.~Hamma,
\newblock \emph{Nonstabilizerness determining the hardness of direct fidelity estimation},
\newblock Phys. Rev. A \textbf{107}, 022429 (2023)
\newblock \doi{10.1103/PhysRevA.107.02242}.

\bibitem{kitaev2014hidden}
A.~Kitaev,
\newblock \emph{Hidden correlations in the {H}awking radiation and thermal
  noise},
\newblock In \emph{Talk given at the Fundamental Physics Prize Symposium,  vol.~10}; \url{https://online.kitp.ucsb.edu/online/joint98/kitaev/}, UC Santa Barbara, (2014).

\bibitem{harrow2021separation}
A.~W. Harrow, L.~Kong, Z.-W. Liu, S.~Mehraban and P.~W. Shor,
\newblock \emph{Separation of out-of-time-ordered correlation and
  entanglement},
\newblock PRX Quantum \textbf{2}, 020339 (2021),
\newblock \doi{10.1103/PRXQuantum.2.020339}.

\bibitem{yang2017entanglement}
Z.-C. Yang, A.~Hamma, S.~M. Giampaolo, E.~R. Mucciolo and C.~Chamon,
\newblock \emph{Entanglement complexity in quantum many-body dynamics,
  thermalization, and localization},
\newblock Phys. Rev. B \textbf{96}, 020408 (2017),
\newblock \doi{10.1103/PhysRevB.96.020408}.

\bibitem{liu2020many}
Z.-W. Liu and A.~Winter,
\newblock \emph{Many-body quantum magic},
\newblock PRX Quantum \textbf{3}, 020333 (2022),
\newblock \doi{10.1103/PRXQuantum.3.020333}.

\bibitem{oliviero2022magic}
S.~F.~E. Oliviero, L.~Leone and A.~Hamma,
\newblock \emph{Magic-state resource theory for the ground state of the
  transverse-field ising model},
\newblock Phys. Rev. A \textbf{106}, 042426 (2022),
\newblock \doi{10.1103/PhysRevA.106.042426}.

\bibitem{oliviero2022measuring}
S.~F.~E. Oliviero, L.~Leone, A.~Hamma and S.~Lloyd,
\newblock \emph{Measuring magic on a quantum processor},
\newblock npj Quantum Information \textbf{8}(1), 1 (2022),
\newblock \doi{10.1038/s41534-022-00666-5}.

\bibitem{haug2023quantifying}
T.~Haug and L.~Piroli,
\newblock \emph{Quantifying nonstabilizerness of matrix product states},
\newblock Phys. Rev. B \textbf{107}, 035148 (2023),
\newblock \doi{10.1103/PhysRevB.107.035148}.

\bibitem{zhou2020single}
S.~Zhou, Z.-C. Yang, A.~Hamma and C.~Chamon,
\newblock \emph{{Single T gate in a Clifford circuit drives transition to
  universal entanglement spectrum statistics}},
\newblock SciPost Phys. \textbf{9}, 87 (2020),
\newblock \doi{10.21468/SciPostPhys.9.6.087}.

\bibitem{oliviero2021transitions}
S.~F. Oliviero, L.~Leone and A.~Hamma,
\newblock \emph{Transitions in entanglement complexity in random quantum
  circuits by measurements},
\newblock Physics Letters A \textbf{418}, 127721 (2021),
\newblock \doi{https://doi.org/10.1016/j.physleta.2021.127721}.

\bibitem{leone2022retrieving}
L.~Leone, S.~F.~E. Oliviero, S.~Piemontese, S.~True and A.~Hamma,
\newblock \emph{Retrieving information from a black hole using quantum machine
  learning},
\newblock Physical Review A \textbf{106}(6), 062434 (2022),
\newblock \doi{10.1103/PhysRevA.106.062434}.

\bibitem{leone2022learning}
L.~Leone, S.~F.~E. Oliviero, S.~Lloyd and A.~Hamma,
\newblock \emph{Learning efficient decoders for quasi-chaotic quantum
  scramblers},
\newblock arXiv,
\newblock \doi{10.48550/arXiv.2212.11338} (2022),
\newblock \href{http://arxiv.org/abs/2212.11338}{{ [quant-ph/2212.11338]}}.

\bibitem{oliviero2022blackhole}
S.~F.~E. Oliviero, L.~Leone, S.~Lloyd and A.~Hamma,
\newblock \emph{Black {{Hole}} complexity, unscrambling, and stabilizer thermal
  machines},
\newblock arXiv,
\newblock \doi{10.48550/arXiv.2212.11337} (2022),
\newblock \href{http://arxiv.org/abs/2212.11337}{{ [gr-qc, physics:hep-th,
  physics:quant-ph/2212.11337]}}.

\bibitem{mahler2005emergence}
G.~Mahler, J.~Gemmer and M.~Michel,
\newblock \emph{Emergence of thermodynamic behavior within composite quantum
  systems},
\newblock Physica E: Low-dimensional Systems and Nanostructures \textbf{29}(1),
  53 (2005),
\newblock \doi{https://doi.org/10.1016/j.physe.2005.05.001},
\newblock Frontiers of Quantum.

\bibitem{popescu2006entanglement}
S.~Popescu, A.~J. Short and A.~Winter,
\newblock \emph{Entanglement and the foundations of statistical mechanics},
\newblock Nature Physics \textbf{2}(11), 754 (2006),
\newblock \doi{10.1038/nphys444}.

\bibitem{goldstein2006canonical}
S.~Goldstein, J.~L. Lebowitz, R.~Tumulka and N.~Zangh\`{\i},
\newblock \emph{Canonical typicality},
\newblock Phys. Rev. Lett. \textbf{96}, 050403 (2006),
\newblock \doi{10.1103/PhysRevLett.96.050403}.

\bibitem{reimann2007typicality}
P.~Reimann,
\newblock \emph{Typicality for generalized microcanonical ensembles},
\newblock Phys. Rev. Lett. \textbf{99}, 160404 (2007),
\newblock \doi{10.1103/PhysRevLett.99.160404}.

\bibitem{rigol2008thermalization}
M.~Rigol, V.~Dunjko and M.~Olshanii,
\newblock \emph{Thermalization and its mechanism for generic isolated quantum
  systems},
\newblock Nature \textbf{452}(7189), 854 (2008),
\newblock \doi{10.1038/nature06838}.



\bibitem{derrida1980rem}
B.~Derrida,
\newblock \emph{Random-energy model: Limit of a family of disordered models},
\newblock Phys. Rev. Lett. \textbf{45}, 79 (1980),
\newblock \doi{10.1103/PhysRevLett.45.79}.

\bibitem{manai2020phase}
C.~Manai and S.~Warzel,
\newblock \emph{Phase diagram of the quantum random energy model},
\newblock Journal of Statistical Physics \textbf{180}(1), 654 (2020),
\newblock \doi{10.1007/s10955-020-02492-5}.

\bibitem{quan2006decay}
H.~T. Quan, Z.~Song, X.~F. Liu, P.~Zanardi and C.~P. Sun,
\newblock \emph{Decay of loschmidt echo enhanced by quantum criticality},
\newblock Phys. Rev. Lett. \textbf{96}, 140604 (2006),
\newblock \doi{10.1103/PhysRevLett.96.140604}.

\bibitem{zanardi2006ground}
P.~Zanardi and N.~Paunkovi\ifmmode~\acute{c}\else \'{c}\fi{},
\newblock \emph{Ground state overlap and quantum phase transitions},
\newblock Phys. Rev. E \textbf{74}, 031123 (2006),
\newblock \doi{10.1103/PhysRevE.74.031123}.

\bibitem{abasto2008fidelity}
D.~F. Abasto, A.~Hamma and P.~Zanardi,
\newblock \emph{Fidelity analysis of topological quantum phase transitions},
\newblock Phys. Rev. A \textbf{78}, 010301 (2008),
\newblock \doi{10.1103/PhysRevA.78.010301}.

\bibitem{castelnovo2005from}
C.~Castelnovo, C.~Chamon, C.~Mudry and P.~Pujol,
\newblock \emph{From quantum mechanics to classical statistical physics:
  Generalized rokhsar–kivelson hamiltonians and the “stochastic matrix
  form” decomposition},
\newblock Annals of Physics \textbf{318}(2), 316 (2005),
\newblock \doi{https://doi.org/10.1016/j.aop.2005.01.006}.

\bibitem{rokhsar1988superconductivity}
D.~S. Rokhsar and S.~A. Kivelson,
\newblock \emph{Superconductivity and the quantum hard-core dimer gas},
\newblock Phys. Rev. Lett. \textbf{61}, 2376 (1988),
\newblock \doi{10.1103/PhysRevLett.61.2376}.

\bibitem{grover2014quantum}
T.~Grover and M.~P.~A. Fisher,
\newblock \emph{Quantum disentangled liquids},
\newblock Journal of Statistical Mechanics: Theory and Experiment
  \textbf{2014}(10), P10010 (2014),
\newblock \doi{10.1088/1742-5468/2014/10/p10010}.

\bibitem{grover2015entanglement}
T.~Grover and M.~P.~A. Fisher,
\newblock \emph{Entanglement and the sign structure of quantum states},
\newblock Phys. Rev. A \textbf{92}, 042308 (2015),
\newblock \doi{10.1103/PhysRevA.92.042308}.

\bibitem{gu2010fidelity}
S.-J. GU,
\newblock \emph{Fidelity approach to quantum phase transitions},
\newblock International Journal of Modern Physics B \textbf{24}(23), 4371
  (2010),
\newblock \doi{10.1142/S0217979210056335},
\newblock \eprint{https://doi.org/10.1142/S0217979210056335}.

\bibitem{zanardi2007information}
P.~Zanardi, P.~Giorda and M.~Cozzini,
\newblock \emph{Information-theoretic differential geometry of quantum phase
  transitions},
\newblock Phys. Rev. Lett. \textbf{99}, 100603 (2007),
\newblock \doi{10.1103/PhysRevLett.99.100603}.

\bibitem{you2007fidelity}
W.-L. You, Y.-W. Li and S.-J. Gu,
\newblock \emph{Fidelity, dynamic structure factor, and susceptibility in
  critical phenomena},
\newblock Phys. Rev. E \textbf{76}, 022101 (2007),
\newblock \doi{10.1103/PhysRevE.76.022101}.

\bibitem{jacobson2009scaling}
N.~T. Jacobson, S.~Garnerone, S.~Haas and P.~Zanardi,
\newblock \emph{Scaling of the fidelity susceptibility in a disordered quantum
  spin chain},
\newblock Phys. Rev. B \textbf{79}, 184427 (2009),
\newblock \doi{10.1103/PhysRevB.79.184427}.

\bibitem{Zhou_2008}
H.-Q. Zhou and J.~P. Barjaktarevi{\v{c}},
\newblock \emph{Fidelity and quantum phase transitions},
\newblock Journal of Physics A: Mathematical and Theoretical \textbf{41}(41),
  412001 (2008),
\newblock \doi{10.1088/1751-8113/41/41/412001}.

\bibitem{Damski_2015}
B.~Damski,
\newblock \emph{Fidelity approach to quantum phase transitions in quantum ising
  model},
\newblock In \emph{Quantum Criticality in Condensed Matter}. {WORLD}
  {SCIENTIFIC},
\newblock \doi{10.1142/9789814704090_0006} (2015).

\bibitem{Zanardietal2008}
P.~Zanardi, M.~G.~A. Paris and L.~Campos~Venuti,
\newblock \emph{Quantum criticality as a resource for quantum estimation},
\newblock Phys. Rev. A \textbf{78}, 042105 (2008),
\newblock \doi{10.1103/PhysRevA.78.042105}.

\bibitem{derrida1980therem}
B.~Derrida,
\newblock \emph{{The random energy model}},
\newblock {Physics Reports} (67), 29 (1980).

%\bibitem{Note1}
%Notice that the critical temperature quoted here differs by a factor of 2 with
  %respect to that of Ref.~\cite {derrida1980rem}, because the variance of
  %energy fluctuations is ${\protect \rm Var}(E_{\protect \bm {\sigma }})= N/2$
  %in that work and it is ${\protect \rm Var}(E_{\protect \bm {\sigma }})= N$ in
  %this work.

\bibitem{newville2014lmfit}
M.~Newville, T.~Stensitzki, D.~B. Allen and A.~Ingargiola,
\newblock \emph{{LMFIT: Non-Linear Least-Square Minimization and Curve-Fitting
  for Python}}, Zenodo (2014),
\newblock \doi{10.5281/zenodo.11813}.

\bibitem{amico2006divergence}
L.~Amico, F.~Baroni, A.~Fubini, D.~Patan\`e, V.~Tognetti and P.~Verrucchi,
\newblock \emph{Divergence of the entanglement range in low-dimensional quantum
  systems},
\newblock Phys. Rev. A \textbf{74}, 022322 (2006),
\newblock \doi{10.1103/PhysRevA.74.022322}.

\bibitem{tomasello2011ground}
B.~Tomasello, D.~Rossini, A.~Hamma and L.~Amico,
\newblock \emph{Ground-state factorization and correlations with broken
  symmetry},
\newblock Europhysics Letters \textbf{96}(2), 27002 (2011),
\newblock \doi{10.1209/0295-5075/96/27002}.

\bibitem{atas2013distribution}
Y.~Y. Atas, E.~Bogomolny, O.~Giraud and G.~Roux,
\newblock \emph{Distribution of the ratio of consecutive level spacings in
  random matrix ensembles},
\newblock Phys. Rev. Lett. \textbf{110}, 084101 (2013),
\newblock \doi{10.1103/PhysRevLett.110.084101}.

\bibitem{Mehta_book}
M.~L. Mehta,
\newblock \emph{Random Matrices},
\newblock Academic Press (2004).

\bibitem{tang2021non-ergodic}
W.~Tang and I.~M. Khaymovich,
\newblock \emph{Non-ergodic delocalized phase with {P}oisson level statistics},
\newblock {Quantum} \textbf{6}, 733 (2022),
\newblock \doi{10.22331/q-2022-06-09-733}.

\bibitem{huse2007localization}
V.~Oganesyan and D.~A. Huse,
\newblock \emph{Localization of interacting fermions at high temperature},
\newblock Phys. Rev. B \textbf{75}, 155111 (2007),
\newblock \doi{10.1103/PhysRevB.75.155111}.

\bibitem{yang2015two-component}
Z.-C. Yang, C.~Chamon, A.~Hamma and E.~R. Mucciolo,
\newblock \emph{Two-component structure in the entanglement spectrum of highly
  excited states},
\newblock Phys. Rev. Lett. \textbf{115}, 267206 (2015),
\newblock \doi{10.1103/PhysRevLett.115.267206}.

\bibitem{virtanen2020scipy}
P.~Virtanen, R.~Gommers, T.~E. Oliphant, M.~Haberland, T.~Reddy, D.~Cournapeau,
  E.~Burovski, P.~Peterson, W.~Weckesser \emph{et~al.},
\newblock \emph{{{SciPy} 1.0: Fundamental Algorithms for Scientific Computing
  in Python}},
\newblock Nature Methods \textbf{17}, 261 (2020),
\newblock \doi{10.1038/s41592-019-0686-2}.

\bibitem{zhu2016clifford}
H.~Zhu, R.~Kueng, M.~Grassl and D.~Gross,
\newblock \emph{The clifford group fails gracefully to be a unitary 4-design},
\newblock \doi{10.48550/ARXIV.1609.08172} (2016).

\end{thebibliography}
	
%\newpage

\end{document}